\documentclass[%
 reprint,
 superscriptaddress,
 amsmath,amssymb,
pra,
]{revtex4-1}

\usepackage{graphicx}
\usepackage{dcolumn}
\usepackage{bm}
\usepackage{braket} 
\usepackage[caption=false]{subfig} 
\usepackage{changepage}
\usepackage{float}
\usepackage{amsmath}
\usepackage{xcolor}
\usepackage{tikz}
\def\checkmark{\tikz\fill[scale=0.4](0,.35) -- (.25,0) -- (1,.7) -- (.25,.15) -- cycle;} 

\begin{document}


\title{On the origin of the Kerker phenomena}

\author{Jon Lasa-Alonso}
\altaffiliation[jonqnanolab@gmail.com]{}
\affiliation{Donostia International Physics Center, Paseo Manuel de Lardizabal 4, 20018 Donostia-San Sebasti\'an, Spain}
\affiliation{Centro de F\'isica de Materiales, Paseo Manuel de Lardizabal 5, 20018 Donostia-San Sebasti\'an, Spain}
\author{Chiara Devescovi}
\affiliation{Donostia International Physics Center, Paseo Manuel de Lardizabal 4, 20018 Donostia-San Sebasti\'an, Spain}
\author{Carlos Maciel-Escudero}
\affiliation{Donostia International Physics Center, Paseo Manuel de Lardizabal 4, 20018 Donostia-San Sebasti\'an, Spain}
\affiliation{Centro de F\'isica de Materiales, Paseo Manuel de Lardizabal 5, 20018 Donostia-San Sebasti\'an, Spain}
\author{Aitzol Garc\'ia-Etxarri}
\affiliation{Donostia International Physics Center, Paseo Manuel de Lardizabal 4, 20018 Donostia-San Sebasti\'an, Spain}
\affiliation{IKERBASQUE, Basque Foundation for Science, Mar\'ia D\'iaz de Haro 3, 48013 Bilbao, Spain}
\author{Gabriel Molina-Terriza}
\affiliation{Donostia International Physics Center, Paseo Manuel de Lardizabal 4, 20018 Donostia-San Sebasti\'an, Spain}
\affiliation{Centro de F\'isica de Materiales, Paseo Manuel de Lardizabal 5, 20018 Donostia-San Sebasti\'an, Spain}
\affiliation{IKERBASQUE, Basque Foundation for Science, Mar\'ia D\'iaz de Haro 3, 48013 Bilbao, Spain}

\date{\today}

\begin{abstract}
We provide an insight into the origin of the phenomena reported 40 years ago by Kerker, Wang and Giles (Journal of the Optical Society of America, 73, 6, pp. 765-767, (1983)). We show that the impedance and refractive index matching conditions, discussed in Sections II and IV of the seminal paper, are intimately related with space-time symmetries. We derive our results starting from the theory of representations of the Poincaré group, as it is the theory on which one of the most elemental descriptions of electromagnetic waves is based. We show that fundamental features of electromagnetic waves in material environments can be derived from group theoretical arguments. In particular, we identify the Casimir invariants of $P_{\scriptscriptstyle{{3,1}}}$ subgroup as the magnitudes which describe the nature of monochromatic electromagnetic waves propagating in matter. Finally, we show that the emergence of the Kerker phenomena is associated with the conservation of such Casimir invariants in piecewise homogeneous media.

\end{abstract}

\maketitle


\section{Introduction} \label{sec:1}
The study of the so-called \emph{Kerker conditions} \cite{Kerker, GilesWild} has gathered an important amount of scientific efforts in the last decade \cite{KerkerApp0, KerkerApp1, KerkerApp2, KerkerRef01, KerkerRef1, Kerker11, KerkerApp4, Kerker13, Kerker14, ACSLasa, Kerker18, SymProt, PRLJorge, Kerker15, Kerker17, Kerker12, Kerker16}. The phrase \emph{first Kerker condition} refers to scatterers that behave as those described in Section II of Kerker's seminal paper \cite{Kerker}. On the other hand, the phrase \emph{second Kerker condition} has been usually employed to make reference to one type of scatterers discussed in Section III. The key characteristics of the scatterers fulfilling either of these two conditions are related to the preservation of helicity and the directionality of the emission. In particular, samples fulfilling Kerker's first condition are characterized for producing a scattered field with the same helicity as the illuminating excitation. As preservation of helicity is related to the restoration of duality symmetry \cite{Calkin}, scatterers fulfilling the first Kerker condition are commonly called dual \cite{PRLMolina}. Scatterers fulfilling Kerker's second condition, however, are characterize for producing a field of the opposite helicity. Consequently, they have usually been denoted as antidual. In addition, in the specific case of rotationally symmetric scatterers, the directionality of the emitted field is directly correlated with its helicity. For example, cylindrical dual scatterers do not emit in the backward direction, whereas cylindrical antidual scatterers do not emit in the forward direction \cite{ZambranaKerker, ForBackCorbaton}.

The features described above have frequently placed the discussion about the two Kerker conditions at the same level. However, we should bear in mind that scatterers fulfilling the first or the second Kerker conditions have contradictory behaviours in many important aspects. For instance, dual scatterers can generally be defined in terms of a condition over the material constants, i.e. they are achieved whenever the impedance of the sample and the surrounding medium is matched. Antidual scatterers, however, have only been identified in terms of multipolar scattering coefficients \cite{ZambranaKerker, Correlations}. Moreover, whereas helicity preserving scatterers have been experimentally reported several times in the literature, antidual scatterers have fundamental problems regarding conservation of energy. These difficulties impulsed the search for alternative definitions of the second Kerker condition that respected the optical theorem. The efforts resulted in the notion of the \emph{generalized second Kerker condition}, that minimizes the forward scattered emission for a fixed scattering cross section \cite{NietoVesperinasK2, PRROlmos}. Even if the generalized second Kerker condition was shown to be compatible with the conservation of electromagnetic energy, it is still a definition based on the Mie scattering coefficients of a sphere. Recently, it has been shown that antidual scatterers of any size and form have, by construction, a null extinction cross section \cite{Resonant}.

The aforementioned differences in the description of dual and antidual scatterers led us to the conclusion that the current description of the two Kerker conditions is not satisfactory. Even if they have been usually discussed on the same footing, it is clear that the nature of these two effects is dissimilar in many important aspects. Here, we propose a solution to this historical conflict with the introduction of the resonant helicity mixing condition. This phenomenon is partially studied in Section IV of Kerker's original paper, but it was not until recently that its crucial role has been put forward \cite{Resonant}. As it was shown therein, the resonant helicity mixing condition is also defined in terms of material constants, i.e. it is fulfilled in samples with matched refractive indices. Moreover, it permits the construction of spherical scatterers which flip helicity of light very efficiently, while still respecting the energy conservation law. Finally, also in line with impedance-matched materials, index-matched materials have an associated conserved quantity: the square of linear momentum. As we show next, the origin of this conserved magnitude lies in the symmetries of space-time and, thus, its comprehension requires a systematic approach to the description of electromagnetic waves in terms of group theory. 

In this work, we first revisit the theory of unitary irreducible representations of the Poincaré group and its link to the description of electromagnetic waves in vacuum. We place at the forefront of our analysis, the important group theoretical concept of Casimir operators, applied to some problems in electromagnetism. In group theory, the Casimir operators of a particular group are operators that commute with all the generators of the group and, thus, label the unitary irreducible representations. In this line, we show that, exactly as Bialynicki-Birula's photon wave function \footnote{In this context, we use the term \emph{photon wave function} as a legacy from Bialynicki-Birula's work. We will always deal in this work with classical electromagnetic fields, and the photon wave functions are electromagnetic wave solutions.} is associated with the unitary irreducible representations of the Poincaré group \cite{IZBirula}, monochromatic electromagnetic waves propagating in infinitely homogeneous media are associated with the unitary irreducible representations of $P_{\scriptscriptstyle{{3,1}}}$ subgroup. As a result, such wave solutions are constructed as eigenfunctions of the three Casimir operators of $P_{\scriptscriptstyle{{3,1}}}$, i.e. the generator of time-translations, $\hat{P}_0$, the helicity operator, $\hat{\Lambda}$, and the square of the linear momentum operator, $\hat{\mathbf{P}}^2$. Finally, we study the propagation of electromagnetic waves in piecewise media and we particularly focus on Kerker's problem, i.e. the scattering of electromagnetic waves by magnetic spheres. We show that the two matching conditions reported in the seminal paper are related with the breaking of the homogeneity of space, while still preserving the Casimir invariants of $P_{\scriptscriptstyle{{3,1}}}$.

\section{Electromagnetic waves in vacuum} \label{sec:2}
One of the most elemental descriptions of the propagation of electromagnetic waves in vacuum is due to Eugene Wigner \cite{Wigner1, BargmannWigner}. His findings not only provided an axiomatic derivation of Maxwell's equations in vacuum, but unified the description of all relativistic particles in terms of the so-called Wigner's classification. His major achievement is probably the connection between  unitary irreducible representations (UIRs) of the Poincaré group and wave functions of isolated physical systems in vacuum. Indeed, he found that the invariant vector spaces of the different UIRs of the Poincaré group are associated with the wave functions of relativistic particles \cite{WuKiTung}. More specifically, Wigner and Bargmann found that the UIRs of the Poincaré group (and, in fact, of many other continuous groups) can be labelled by the eigenvalues of the two principal Casimir operators: $\hat{C}_1 = \hat{P}_\nu \hat{P}^\nu$ and $\hat{C}_2 = \hat{W}_\nu \hat{W}^\nu$, which are the modulus of the $4$-momentum, $\hat{P}_\nu$, and the modulus of the Pauli-Lubanski pseudovector, $\hat{W}_\nu$, respectively \cite{BargmannWigner, Bargmann}. In practice, the first Casimir operator represents the mass, whereas the second Casimir operator is associated with the internal degrees of freedom of the particles. In this Section, we will mostly focus on electromagnetic waves propagating in vacuum and, thus, we will center our analysis on the massless and discrete-spin UIRs of the Poincaré group. For this type of particles, a third Casimir operator is identified \cite{ZeroMassRep}: helicity, $\hat{\Lambda}$.

The description of the photon, as a fundamental relativistic particle, is considered within the $0_s$ class of UIRs of the Poincaré group \cite{BargmannWigner}. This class describes particles of null mass, $\text{\emph{M}} = 0$, and helicity eigenvalue $\lambda = \pm \text{\emph{S}}$, where \emph{S} is an integer or an odd-half-integer \cite{WuKiTung}. The electromagnetic case is recovered when fixing $\text{\emph{S}} = 1$. The invariant vector spaces associated with the description of the photon fulfill the following relations: $\hat{C}_1 = 0$ and $\hat{W}_\nu = \lambda \hat{P}_\nu$ \cite{BargmannWigner}. Even if the form might be cumbersome, it can be shown that the first equation represents the wave equation in vacuum. Moreover, the time component of the second equation represents Faraday-Ampère's laws in vacuum and, finally, Gauss' laws can be derived from the spatial components of the second equation \cite{HBacry, Gersten, Dirac}. In our view, the deepest explicit connection between the UIRs of the Poincaré group and Maxwell's equations has been carried out by Iwo Bialynicki-Birula and Zofia Bialynicka-Birula. In their works, they argue that the electromagnetic object which more closely follows the UIRs of the Poincaré group is the Riemann-Silberstein (RS) vector, from which a proper photon wave function can be constructed. Explicitly, they have proposed the following form of the photon wave function in vacuum \cite{IZBirula}:
\begin{equation}
    \label{PWF}
    \mathbf{\Psi}^\lambda(\mathbf{r}, t) = \frac{1}{(2\pi)^{3/2}}\int d\mathbf{k}~f^\lambda(\mathbf{k})\mathbf{e}^\lambda(\mathbf{k})e^{i(\mathbf{k}\cdot\mathbf{r} - \omega_\mathbf{k} t)},
\end{equation}
where $\mathbf{r}$ is the position vector and $t$ the time. On the other hand, $\lambda = \pm 1$ is the helicity label, $\mathbf{k}$ is the wavevector of the radiation field, $f^\lambda(\mathbf{k})$ is an arbitrary complex amplitude associated with a fixed wavevector and helicity, $\mathbf{e}^\lambda(\mathbf{k})$ is a unitary polarization vector and $\omega_\mathbf{k} = |\mathbf{k}|$ is the angular frequency (we choose natural units, $\hbar = c = 1$). Finally, the integral in Eq. \eqref{PWF} is considered over the entire reciprocal space.

The expression in Eq. \eqref{PWF} of the photon wave function has been derived as a particular solution of Maxwell's equations in terms of the RS vector. However, let us now show that Eq. \eqref{PWF} can also be derived by constructing the invariant vector spaces of the massless UIRs of the Poincaré group. This connection highlights that important features of electromagnetic wave solutions in vacuum can be derived from pure group theoretical arguments. In this regard, we closely follow the method indicated by Wu-Ki Tung's book, \emph{Group Theory in Physics} \cite{WuKiTung}, which indicates that the construction of invariant spaces can be carried out by operating over a ``standard" vector. Indeed, the basis vectors of the massless and discrete-spin UIRs of the Poincaré group are constructed as (see Ref. \cite{WuKiTung}, Chapter 10, Section 4):
\begin{equation}
    \label{UIR_Poincare}
   \boldsymbol{\Psi}^\lambda_{\mathbf{k}}(\mathbf{r}, t) \equiv \left[\hat{R}_z(\phi)\hat{R}_y(\theta)\hat{L}_z(\xi)\right]\boldsymbol{\Psi}^\lambda_{\mathbf{k}_l}(\mathbf{r}, t).
\end{equation}
where $\boldsymbol{\Psi}^\lambda_{\mathbf{k}_l}(\mathbf{r}, t) = \mathbf{u}^\lambda e^{ik_l(z - t)}$ represents a monochromatic plane wave propagating in the positive $OZ$ direction with linear momentum $\mathbf{k}_l$ and $\mathbf{u}^\lambda = (1, \lambda i, 0)/\sqrt{2}$ is a polarization vector with well-defined helicity $\lambda$. $\hat{R}_z(\phi)$ represents a rotation along $OZ$ axis by an angle $\phi$ and $\hat{R}_y(\theta)$ a rotation along $OY$ axis by an angle $\theta$. $\hat{L}_z(\xi)$ represents a Lorentz transformation along the $OZ$ direction to an inertial frame moving with velocity $v = \tanh(\xi)$. It can be checked that the application of the Lorentz transformation modifies the frequency of the standard vector as follows: $|\mathbf{k}| = e^{-\xi}|\mathbf{k}_l|$. Thus, by choosing the boost parameter $\xi \in (-\infty, \infty)$ one spans all possible values of the wavevector modulus (see Appendix \ref{AppendixA}). On the other hand, by choosing $\phi \in (0, 2\pi)$ and $\theta \in (0,\pi)$ the rotation transformation makes the monochromatic plane wave propagate in an arbitrary direction.

If we now consider the integral in Eq. \eqref{PWF} in spherical coordinates of reciprocal space, we see that the three parameters $\{\xi, \theta, \phi\}$ span all the integration domain. In other words, the photon wave function in vacuum given in Eq. \eqref{PWF} can be written in the following form:
\begin{equation}
    \label{PWF_WuKiTung}
    \mathbf{\Psi}^\lambda(\mathbf{r}, t) = \frac{1}{(2\pi)^{3/2}} \int d\mathbf{k}~g^\lambda(\mathbf{k}) \boldsymbol{\Psi}^\lambda_{\mathbf{k}}(\mathbf{r}, t),
\end{equation}
where the introduction of a different complex function, $g^\lambda(\mathbf{k}) = e^{\xi} f^\lambda(\mathbf{k})$, is just because the Lorentz transformation not only changes the frequency, but also modifies the amplitude of the standard vector by a factor $e^{-\xi}$ (see Appendix A). Also note that, even if the construction in Eq. \eqref{UIR_Poincare} starts from a monochromatic wave, the final state in Eq. \eqref{PWF_WuKiTung} is no longer monochromatic. From this result we conclude that one can retrieve the expression of the photon wave function in vacuum, in the form proposed by Bialynicki-Birula, just by taking linear superpositions of the basis vectors of the massless and discrete-spin UIRs of the Poincaré group. Now, the question arises: following similar group theoretical arguments, can we study the propagation of electromagnetic waves in a different medium other than vacuum?


\section{Electromagnetic waves in infinitely homogeneous media} \label{sec:3}

Let us first consider the next simplest case, i.e. a medium where both the electric permittivity, $\varepsilon$, and magnetic permeability, $\mu$, are different from the values of vacuum but are constant functions of space coordinates. If such a medium extends all over the space, we say that it is infinitely homogeneous. Moreover, we will also consider that the permittivity and permeability are scalar functions (isotropic) and that they do not change on time (static).

\begin{figure}[t]
    \centering
    \includegraphics[width = 0.45\textwidth]{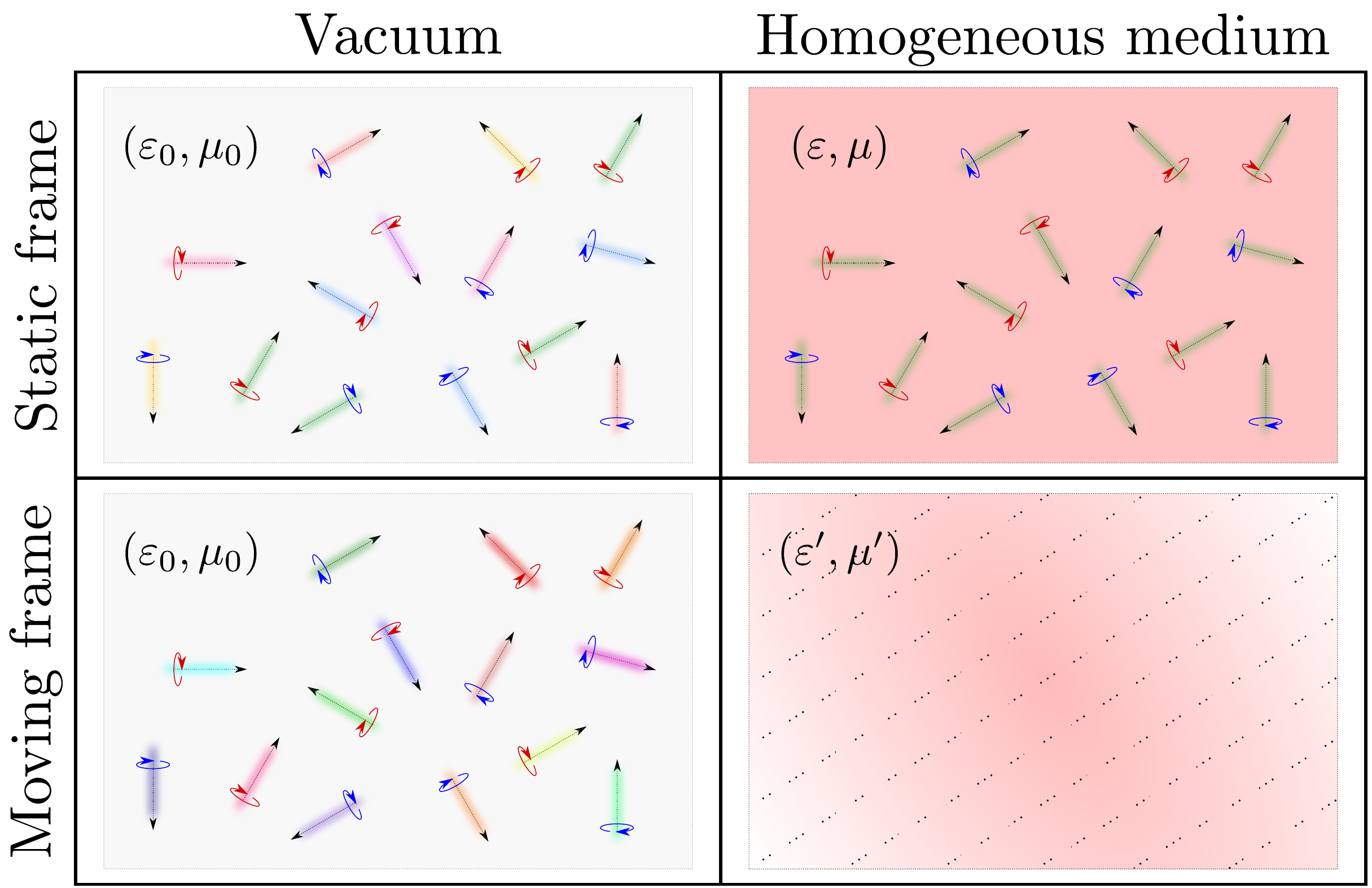}
    \caption{Effect of a Lorentz transformation over electromagnetic waves propagating in vacuum and in a homogeneous medium. Black arrows represent the propagation directions, $(\theta,\phi)$, red/blue spinning arrows the helicity, $\lambda$, and different colours indicate the frequency of the waves, $\omega$. In the left upper panel, a superposition of plane-waves of different frequencies propagating in vacuum is shown. When switching to the moving reference frame (left bottom panel), the environment is left unaltered and the waves change their frequency. In the right upper panel, a superposition of monochromatic plane-waves is shown. When switching to the moving reference frame (right bottom panel), the medium is modified and becomes bianisotropic.}
    \label{VacuumWaves}
\end{figure}

The key modification of the problem when studying the propagation of electromagnetic waves in a infinitely homogeneous medium compared to the case of vacuum is the set of underlying symmetries. An infinitely homogeneous medium is not invariant under the Poincaré group. This is due to the fact that an isotropic medium becomes, in general, bianisotropic when performing a Lorentz transformation (see Fig. \ref{VacuumWaves}) \cite{Landau, BiTheorem}. As a result, the presence of a material reduces the symmetry group from the Poincaré group to a subgroup which does not contain Lorentz transformations. Such a particular way of addressing physical problems is commonly denoted as the symmetry breaking principle
and it can be compactly stated in the following terms \cite{SymmetryBreak1, SymmetryBreak2, SBGeneral1, SBGeneral2, SBGeneral3, SymmetryBreak3, SymmetryBreak4}: \emph{consider a physical system which is described by a given group $G$ and an external influence reduces the symmetry from the original $G$ to a subgroup $G_i \subset G$. Then, the subgroup $G_i$ can be used to study the properties of the new modified system. In particular, the generators and Casimir operators of the subgroup will provide conserved quantities and the UIRs will determine the new wave functions or, at least, some of their properties.}

So, let us apply the symmetry breaking principle to study the propagation of electromagnetic waves in an infinitely homogeneous medium. As stated above such a medium is not invariant under the full Poincaré group, but only under the subgroup $P_{\scriptscriptstyle{{3,1}}}$ which includes all the generators of the Poincaré group except for Lorentz boosts. In other words, $P_{\scriptscriptstyle{{3,1}}}$ comprises seven generators: the generator of time translations ($\hat{P}_0$), the three generators of space translations (linear momentum components, $\hat{P}_i$) and the three generators of spatial rotations (total angular momentum components, $\hat{J}_i$). In addition to this, the subgroup has three Casimir operators, i.e. the generator of time translations, $\hat{P}_0$, a magnitude proportional to the helicity, $\hat{\mathbf{J}}\cdot\hat{\mathbf{P}}$, and the square of linear momentum, $\hat{\mathbf{P}}^2$ \cite{SymmetryBreak4}. Helicity and square of linear momentum are also conserved magnitudes for electromagnetic waves propagating in infinitely homogeneous media due to their condition of Casimir invariants. These invariants will be of utmost importance in Sections IV and V, when discussing the conserved quantities associated with the Kerker phenomena.

Casimir operators play a central role in the determination of the UIRs. This is because the basis vectors of the UIRs are necessarily eigenvectors of all such operators \cite{BargmannWigner, Bargmann, WuKiTung}. Thus, for the basis vectors associated with the UIRs of $P_{\scriptscriptstyle{{3,1}}}$ subgroup, $\boldsymbol{\Phi}^\lambda_{\mathbf{k}}(\mathbf{r}, t)$, we have that:
\begin{align}
    \label{CasimirH}
    \hat{P}_0~\boldsymbol{\Phi}^\lambda_{\mathbf{k}}(\mathbf{r}, t) &= \omega \boldsymbol{\Phi}^\lambda_{\mathbf{k}}(\mathbf{r}, t)\\
    \label{CasimirJP}
    k^{\scriptscriptstyle{-1}}\hat{\mathbf{J}}\cdot\hat{\mathbf{P}}~\boldsymbol{\Phi}^\lambda_{\mathbf{k}}(\mathbf{r}, t) &= \lambda \boldsymbol{\Phi}^\lambda_{\mathbf{k}}(\mathbf{r}, t)\\
    \label{CasimirP2}
    \hat{\mathbf{P}}^2~\boldsymbol{\Phi}^\lambda_{\mathbf{k}}(\mathbf{r}, t) &= k^2 \boldsymbol{\Phi}^\lambda_{\mathbf{k}}(\mathbf{r}, t).
\end{align}
Setting $\lambda = \pm 1$ and $k = \omega n$, with $n = \sqrt{\varepsilon\mu}$ the refractive index of the medium, it can be noted that Eqs. \eqref{CasimirH}-\eqref{CasimirP2} actually represent the dynamic equations of monochromatic electromagnetic waves propagating in an infinitely homogeneous medium. Indeed, Eq. \eqref{CasimirH} determines the monochromaticity of the fields; on the other hand, Eq. \eqref{CasimirJP} represents all four time-independent Maxwell's equations; and, finally, Eq. \eqref{CasimirP2} represents the wave equation for monochromatic  waves, i.e. Helmholtz's equation. The connection appears more clearly when employing the explicit form of the operators \cite{BliokhOperators, PhysRevA.86.042103, HalasHilbert, PhysRevA_Hilbert}: $\hat{P}_0 = i\partial_t$, $\hat{\Lambda} = k^{\scriptscriptstyle{-1}}\hat{\mathbf{J}}\cdot\hat{\mathbf{P}} = k^{\scriptscriptstyle{-1}}\nabla \times$ and $\hat{\mathbf{P}}^2 = -\nabla^2$. This is in agreement with the results previously obtained for vacuum, where Casimir operators were also related with the dynamic equations of electromagnetic waves.

At this stage, following the same procedure as in the case of vacuum, we should be able to obtain some information of the photon wave function in an infinitely homogeneous medium by analyzing the UIRs of $P_{\scriptscriptstyle{{3,1}}}$. Particularly, $P_{\scriptscriptstyle{{3,1}}}$ can be split as the direct product of two of its subgroups which are the Euclidean group in three dimensions, $E(3)$, and the one parameter subgroup of time translations, $T$. This is due to the fact that time translations commute with all the elements of the Euclidean group in three dimensions. As a result, the UIRs of $P_{\scriptscriptstyle{{3,1}}}$ are constructed as the product of the UIRs of $E(3)$ and the UIRs of $T$ \cite{Mackey, Rossmann}. An intuitive way of understanding this is by noting that the basis vectors $\boldsymbol{\Phi}^\lambda_{\mathbf{k}}(\mathbf{r}, t)$ are necessarily built as the product of a spatial function and a temporal function. Equations \eqref{CasimirJP}-\eqref{CasimirP2} determine the spatial dependence of the vectors, whereas Eq. \eqref{CasimirH} determines their temporal dependence. Thus, the basis vectors of the UIRs of $P_{\scriptscriptstyle{{3,1}}}$ associated with electromagnetic waves can be constructed in the following way (see Ref. \cite{WuKiTung}, Chapter 9, Section 7):
\begin{equation}
    \label{UIR_P31}
    \boldsymbol{\Phi}^\lambda_{\mathbf{k}}(\mathbf{r}, t) \equiv \left[\hat{R}_z(\phi)\hat{R}_y(\theta)\right] \boldsymbol{\Phi}^\lambda_{\mathbf{k}_0}(\mathbf{r})e^{-i\omega t}.
\end{equation}
where $\boldsymbol{\Phi}^\lambda_{\mathbf{k}_0}(\mathbf{r}) = \mathbf{u}^\lambda e^{i k z}$ is now the standard vector and $\mathbf{u}^\lambda = (1, \lambda i, 0)/\sqrt{2}$ is a circular polarization vector.

Note that the basis given in Eq. \eqref{UIR_P31} bears a close resemblance to the basis previously constructed for the UIRs of the Poincaré group. Indeed, they are constructed exactly in the same way except for the Lorentz boost applied in Eq. \eqref{UIR_Poincare}, which permits the modulation of the frequency through parameter $\xi$. The basis vectors given by Eq. \eqref{UIR_P31}, on the other hand, have a fixed frequency $\omega$ (see Table \ref{table}). As a result, we can conclude that the photon wave function in an infinitely homogeneous medium can be written as:
\begin{equation}
    \label{PWF_medium}
    \mathbf{\Phi}^\lambda(\mathbf{r}, t) = \int d\Omega~\varphi^\lambda(\Omega)\boldsymbol{\Phi}^\lambda_\mathbf{k}(\mathbf{r}, t),
\end{equation}
where $d\Omega = \sin\theta d\theta d\phi$, with integration limits $\phi \in (0,2\pi)$ and $\theta \in (0,\pi)$, and $\varphi^\lambda(\Omega)$ is an arbitrary complex function of the azimuthal and polar angles. Note that such a field is nothing but an abstract representation of the monochromatic RS vector. As a matter of fact, such an expression of the photon wave function in an infinitely homogeneous medium has already been employed to construct solutions of electromagnetic waves with well-defined helicity, in particular, vector spherical harmonics and Bessel beams \cite{WuKiTung,PhysRevA.86.042103,PolychromaticTmatrix}. The way in which we have derived it here indicates that the validity of the expression lies in the underlying symmetry group and that it goes far beyond the specific examples previously reported. Our findings indicate that the role of the monochromatic RS vector is central in the description of optical phenomena in material environments.
\begin{table}[b]
\begin{center}
\begin{tabular}{ c|c|c|c|c|c|c }
 & \hspace{0.2cm}{$\hat{C}_0$}\hspace{0.2cm} & \hspace{0.2cm}{$\hat{C}_1$}\hspace{0.2cm} & \hspace{0.2cm}{$\hat{C}_2$}\hspace{0.2cm} & \hspace{0.25cm}{$\hat{\Lambda}$}\hspace{0.25cm} & \hspace{0.2cm}{$\hat{P}_0$}\hspace{0.2cm} & \hspace{0.2cm}{$\hat{\mathbf{P}}^2$}\hspace{0.2cm}\\
\hline
& & & & & &\\
{\hspace{0.2cm}$\boldsymbol{\Psi}^\lambda(\mathbf{r}, t)$}\hspace{0.2cm} & \checkmark & \checkmark & \checkmark & \checkmark & &  \\
& & & & & &\\
{\hspace{0.2cm}$\boldsymbol{\Phi}^\lambda(\mathbf{r}, t)$}\hspace{0.2cm} & \checkmark & \checkmark & & \checkmark & \checkmark & \checkmark \\
& & & & & &

\end{tabular}
\caption{Behaviour of the photon wave functions in vacuum, $\boldsymbol{\Psi}^\lambda(\mathbf{r}, t)$, and in a homogeneous medium, $\boldsymbol{\Phi}^\lambda(\mathbf{r}, t)$, under different Casimir operators. The check mark indicates that the wave function is an eigenvector of the indicated Casimir operator. $\hat{C}_0 = \hat{P}_0/|\hat{P}_0|$ is the sign of frequency, $\hat{C}_1 = \hat{\mathbf{P}}^2 - \hat{P}_0^2$ is the modulus of the 4-momentum and $\hat{C}_2 = \hat{\mathbf{W}}^2 - (\hat{\mathbf{J}}\cdot\hat{\mathbf{P}})^2$ the modulus of the Pauli-Lubanski pseudo-vector. $\hat{\mathbf{W}} = \hat{P}_0\hat{\mathbf{J}} + \hat{\mathbf{K}}\times\hat{\mathbf{P}}$ is the spatial component of the Pauli-Lubanski pseudovector, with $\hat{\mathbf{K}}$ the boost operator, i.e. the generator of Lorentz transformations. Finally, note that $\hat{P}_0$ and $\hat{\mathbf{P}}^2$ operators label different UIRs of the $P_{\scriptscriptstyle 3,1}$ subgroup, whereas Poincaré UIRs are superpositions of monochromatic waves with different frequencies.}
\label{table}
\end{center}
\end{table}
\noindent

All the analysis above indicates that the dynamic equations that are employed to study the propagation of electromagnetic waves in infinitely homogeneous media can be derived from pure group theoretical arguments. And, in particular, it highlights the fundamental role that the Casimir operators $\hat{P}_0$, $\hat{\Lambda}$ and $\hat{\mathbf{P}}^2$ play in the description of electromagnetic waves propagating in material environments. Also, note that the symmetry breaking principle had previously been applied for non-relativistic massive particles \cite{SymmetryBreak1, SymmetryBreak3}. Indeed, the time-independent Schrödinger's equation in an infinitely homogeneous medium can also be associated with Eqs. \eqref{CasimirH}-\eqref{CasimirP2} simply by fixing $\lambda = 0$ and $k = \sqrt{2m(E-V)}$, where $E$ is the eigenvalue of $\hat{P}_0$ operator and $V$ represents a constant potential \cite{SymmetryBreak1, SymmetryBreak3, SymmetryBreak5, Mackey}. This mathematical analogy between the dynamic equations and wave functions describing massless and massive particles will aid us later in the discussion of the Kerker phenomena.

\section{Electromagnetic waves in piecewise media: impedance and refractive index matching} \label{sec:4}

\begin{figure}[t]
    \centering
    \includegraphics[width = 0.4\textwidth]{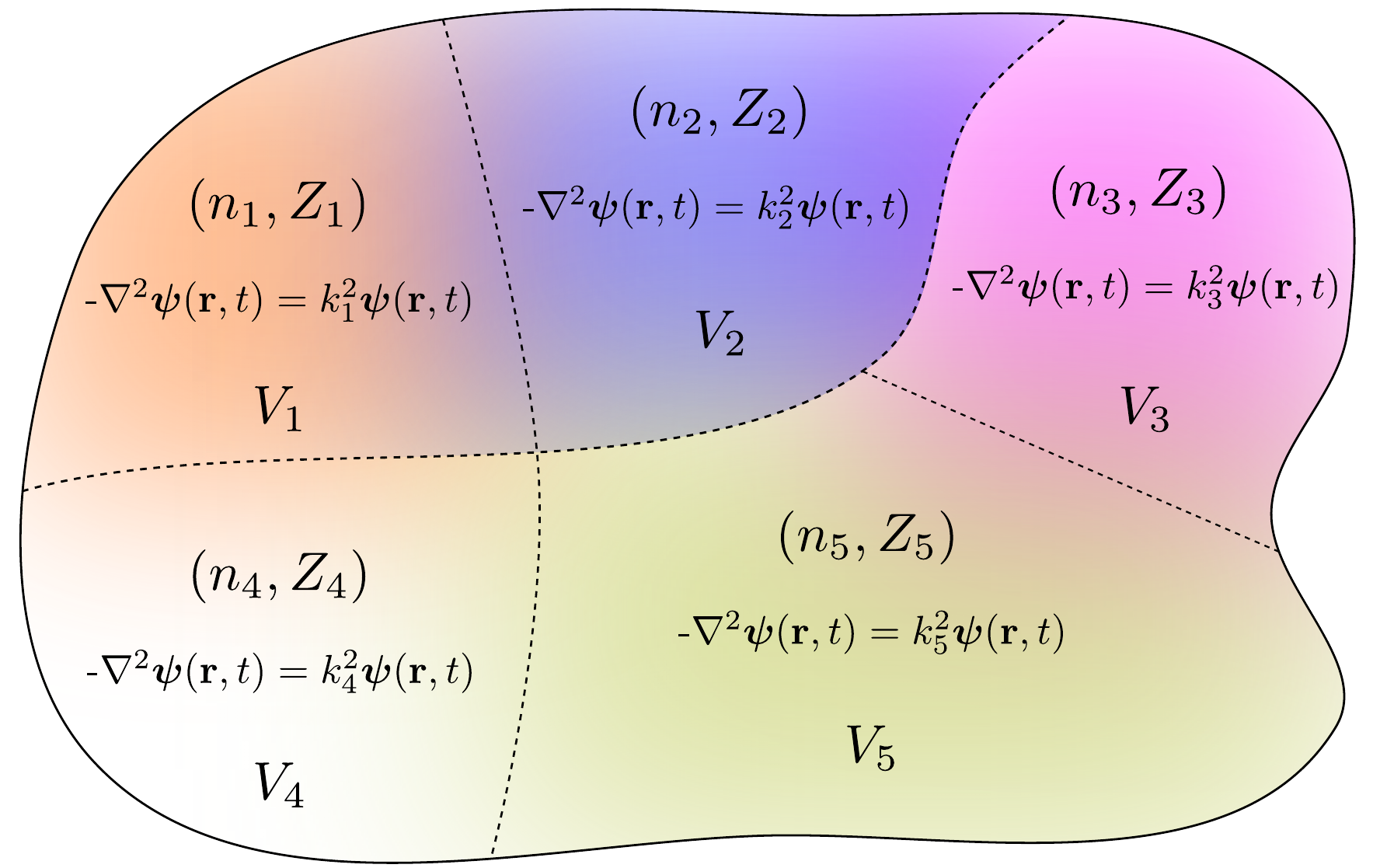}
    \caption{Sketch of a piecewise homogeneous medium. Electromagnetic wave solutions are obtained by solving Helmholtz's equation and applying boundary conditions at the interfaces.}
    \label{PiecewiseHomo}
\end{figure}

In static piecewise homogeneous media there is only one symmetry left, i.e. the one\textcolor{blue}{-}parameter subgroup of time translations, $T$. It is so because this type of environments are not invariant under spatial translations and rotations in three dimensions (see Fig. \ref{PiecewiseHomo}). As a result, the only symmetry of the system is due to the static nature of the medium.

Only a few things can be generally stated about the dynamics in this type of media, i.e. that frequency, $\omega$, is conserved and that the electromagnetic wave solutions are eigenstates of $\hat{P}_0$. Indeed, this result is extensively employed to solve problems of very different nature in physics such as time-independent electromagnetic or quantum scattering problems. The UIRs of the group of time translations can be labelled as (see Ref. \cite{WuKiTung}, Chapter 6, Section 6):
\begin{equation}
    \label{UIR_T}
    \hat{P}_0\boldsymbol{\psi}(\mathbf{r}, t) = \omega \boldsymbol{\psi}(\mathbf{r}, t).
\end{equation}
This condition is equivalent to Eq. \eqref{CasimirH}. As before, it implies that the time dependence of the solutions in piecewise homogeneous media is fixed and it is given by a complex exponential. On the contrary, the spatial functional form of the solutions will strongly depend on the specific symmetries of the particular system one may consider. The generators of the subgroups of $E(3)$ are not the same and, thus, solutions in piecewise media are, in general, eigenfunctions of different sets of operators \cite{SymmetryBreak3, SymmetryBreak4}. As a result, conserved quantities are expected to be strongly dependent on the particular geometries.

In addition, there exist two situations under which electromagnetic fields have an unusual behaviour in piecewise homogeneous media. Such conditions can be identified from the form that Maxwell's equations acquire in terms of the monochromatic RS vector, $\boldsymbol{\Phi}^\lambda(\mathbf{r}, t)$. Indeed, in a generic inhomogeneous medium Maxwell's equations can be expressed as \cite{Birula1, Resonant}:
\begin{align}
    \label{Max_InRS1}
    i\partial_t\boldsymbol{\Phi}^+ &= \frac{1}{\sqrt{n}}\nabla\times\left(\frac{\boldsymbol{\Phi}^+}{\sqrt{n}}\right) + \frac{1}{n}\nabla\ln\sqrt{Z}\times\boldsymbol{\Phi}^-\\
    \label{Max_InRS2}
    i\partial_t\boldsymbol{\Phi}^- &= -\frac{1}{\sqrt{n}}\nabla\times\left(\frac{\boldsymbol{\Phi}^-}{\sqrt{n}}\right) - \frac{1}{n}\nabla\ln\sqrt{Z}\times\boldsymbol{\Phi}^+,
\end{align}
where, for convenience, we have omitted the dependence on the position vector and time. Also, we have defined the local impedance $Z(\mathbf{r}) = \sqrt{\mu(\mathbf{r})/\varepsilon(\mathbf{r})}$ and local refractive index $n(\mathbf{r}) = \sqrt{\varepsilon(\mathbf{r})\mu(\mathbf{r})}$. Given the form of Maxwell's equations in Eqs. \eqref{Max_InRS1}-\eqref{Max_InRS2}, we can identify two specific types of environments in which the electromagnetic field may have a particular behaviour: media for which the impedance is constant ($\nabla Z = 0$) or environments for which the refractive index is constant ($\nabla n = 0$). We usually denote these two situations as the impedance matching and the refractive index matching conditions, respectively, and they are discussed in Sections II and IV of Kerker's paper \cite{Kerker}. Note that, in piecewise media, the impedance matching condition is fulfilled when $Z_j = \sqrt{\mu_j/\varepsilon_j}$ is constant in all domains $V_j$. On the other hand, the refractive index matching condition is fulfilled whenever $n_j = \sqrt{\varepsilon_j\mu_j}$ is constant in all domains $V_j$.

In this line, Fernandez-Corbaton and coworkers indicated that the impedance matching condition leads to the conservation of electromagnetic helicity \cite{PRLMolina}. It was shown that, whenever $\mu_j/\varepsilon_j$ is constant, Maxwell's equations in the whole piecewise medium remain invariant under electromagnetic duality transformations. As helicity had previously been identified as the generator of the duality transformation \cite{Calkin}, it was then concluded that helicity had to be preserved in an impedance-matched piecewise homogeneous medium. However, there is also a dynamical way of understanding the conservation of helicity in impedance-matched media. Indeed, given the form of Maxwell's equations specified in Eqs. \eqref{Max_InRS1}-\eqref{Max_InRS2}, it can be checked that the impedance matching condition makes the two helicity components of the electromagnetic field, $\boldsymbol{\Phi}^+(\mathbf{r}, t)$ and $\boldsymbol{\Phi}^-(\mathbf{r}, t)$ to be decoupled. This implies that, if we fix the initial conditions at time $t = 0$ to contain a single helicity component, the solutions in an impedance-matched piecewise medium will have a single helicity component at times $t > 0$. In the case of piecewise media, this implies that the solutions under the impedance matching conditions fulfill: $\hat{\Lambda}\boldsymbol{\psi}_\lambda(\mathbf{r}, t) = \lambda\boldsymbol{\psi}_\lambda(\mathbf{r}, t)$, with $\lambda = \pm1$. This means that we can also infer the conservation of helicity in piecewise media from the fact that eigenstates of $\hat{\Lambda}$, acting as the Casimir operator in Eq. \eqref{CasimirJP}, remain eigenstates of $\hat{\Lambda}$ in dual media. 

In this sense, the helicity operator has a double role. On the one hand is the generator of dual transformations, but as seen in Eq. \eqref{CasimirJP}, it also acts as a Casimir operator. This could bring the question of whether Casimir operators are fundamentally interesting to study electromagnetic wave dynamics in piecewise homogeneous media. We will show now that in the case of the refractive index matching condition, the conserved quantity is not associated with the generator of a transformation, but with another Casimir operator: $\hat{\mathbf{P}}^2$. In this case, the conservation of a physical magnitude can be inferred from how electromagnetic wave solutions are built in piecewise media. Indeed, as it is shown in Fig. \ref{PiecewiseHomo}, solutions of electromagnetic waves propagating in this type of environments are constructed by solving Helmholtz's equation in each domain $V_j$ and, then, applying boundary conditions. This implies that, in a generic piecewise homogeneous medium, electromagnetic wave solutions fulfill, $-\nabla^2\boldsymbol{\psi}(\mathbf{r}, t) = k_j^2\boldsymbol{\psi}(\mathbf{r}, t)$ for $\mathbf{r}\in V_j$, where $k_j = \omega n_j$. Note that, for a generic piecewise homogeneous medium, $k_j$ changes depending on the region of space we may consider. Thus, the eigenvalue of $\hat{\mathbf{P}}^2 = -\nabla^2$ operator varies from one region $V_j$ to another $V_{j'}$, as long as $n_j \neq n_{j'}$. However, whenever the refractive index matching condition is fulfilled ($n_j = n_{j'},~\forall j,j'$), the wavevector modulus is constant all over the medium and the square of linear momentum operator fulfills: $\hat{\mathbf{P}}^2\boldsymbol{\psi}_k(\mathbf{r}, t) = k^2\boldsymbol{\psi}_k(\mathbf{r}, t)$, where now $k$ is fixed. This is exactly the way in which the conservation of square of linear momentum is represented in Eq. \eqref{CasimirP2}. In this line, the refractive index matching condition recovers that same relation, but in piecewise environments.

%

Following the previous discussion on helicity conservation, we may try to link the conservation of $\hat{\mathbf{P}}^2$ with the restoration of a symmetry. This, however, is not possible because the square of linear momentum is not a generator of any continuous symmetry transformation. Therefore, we are led to comprehend the conservation of $\hat{\mathbf{P}}^2$ from its condition of Casimir invariant. Indeed, Casimir operators of a group \emph{G} may remain as conserved quantities in environments whose symmetry group is a subgroup \emph{G$_i \subset $ G} \cite{SubgroupsPoincare}. This phenomenon is not a particularity of electromagnetic waves, it is known to occur also in the dynamics of massive particles. Let us put forward a particular example which very closely mimics the refractive index matching condition for electromagnetic waves. As we have shown in Section \ref{sec:3}, the conservation of $\hat{\mathbf{P}}^2$ for non-relativistic massive particles can also be related to the symmetries of Euclidean space \cite{Mackey}. Thus, in principle, we may expect the square of linear momentum to be exclusively preserved when massive particles propagate in free space. This, however, is not true. There is a particular interaction potential, i.e. the hard sphere potential, for which $\hat{\mathbf{P}}^2$ is a conserved quantity (see Appendix \ref{AppendixB}). The same also holds for classical mechanics, where the kinetic energy, proportional to the square of the linear momentum, is known to be preserved provided that the interactions are elastic. To the best of our knowledge, the conservation of kinetic energy in elastic collisions has not been linked before with the restoration of a space-time symmetry.

There are more cases in which a Casimir invariant of a group \emph{G} is preserved in an environment whose symmetry group is a subgroup \emph{G$_i$}. A well-known example is the conservation of mass \emph{M} and spin \emph{S}. As we have shown in Section \ref{sec:2}, the conservation of these magnitudes can be concluded from their condition of Casimir invariants of the Poincaré group. This, in principle, would imply that \emph{M} and \emph{S} are exclusively preserved for relativistic massive particles propagating in vacuum. We know, however, that these magnitudes are preserved in a great deal of problems for which the symmetry group is a subgroup of the Poincaré group. For instance, there are many situations in which the mass and spin of the particles are preserved in the framework of non-relativistic quantum scattering theory. Indeed, mass is quite generally assumed to be conserved in this context, whereas spin may not be conserved only in particular cases such as spin-orbit or nucleon-nucleon interactions \cite{ScatteringTaylor}. All these examples point in the direction that the conservation of certain physical magnitudes cannot be related with the restoration of a symmetry, but to their condition of Casimir invariants.

\section{Origin of the Kerker phenomena} \label{sec:5}
Finally, let us apply the previous analysis on conserved quantities to a case of special significance, i.e. the emergence of the Kerker phenomena in magnetic spheres. To be clear in the wording, instead of the usual phrase \emph{Kerker conditions}, we employ \emph{Kerker phenomena} to denote both the impedance and refractive index matching conditions in magnetic spheres. 
\begin{figure}[t]
    \centering
    \includegraphics[width = 0.45\textwidth]{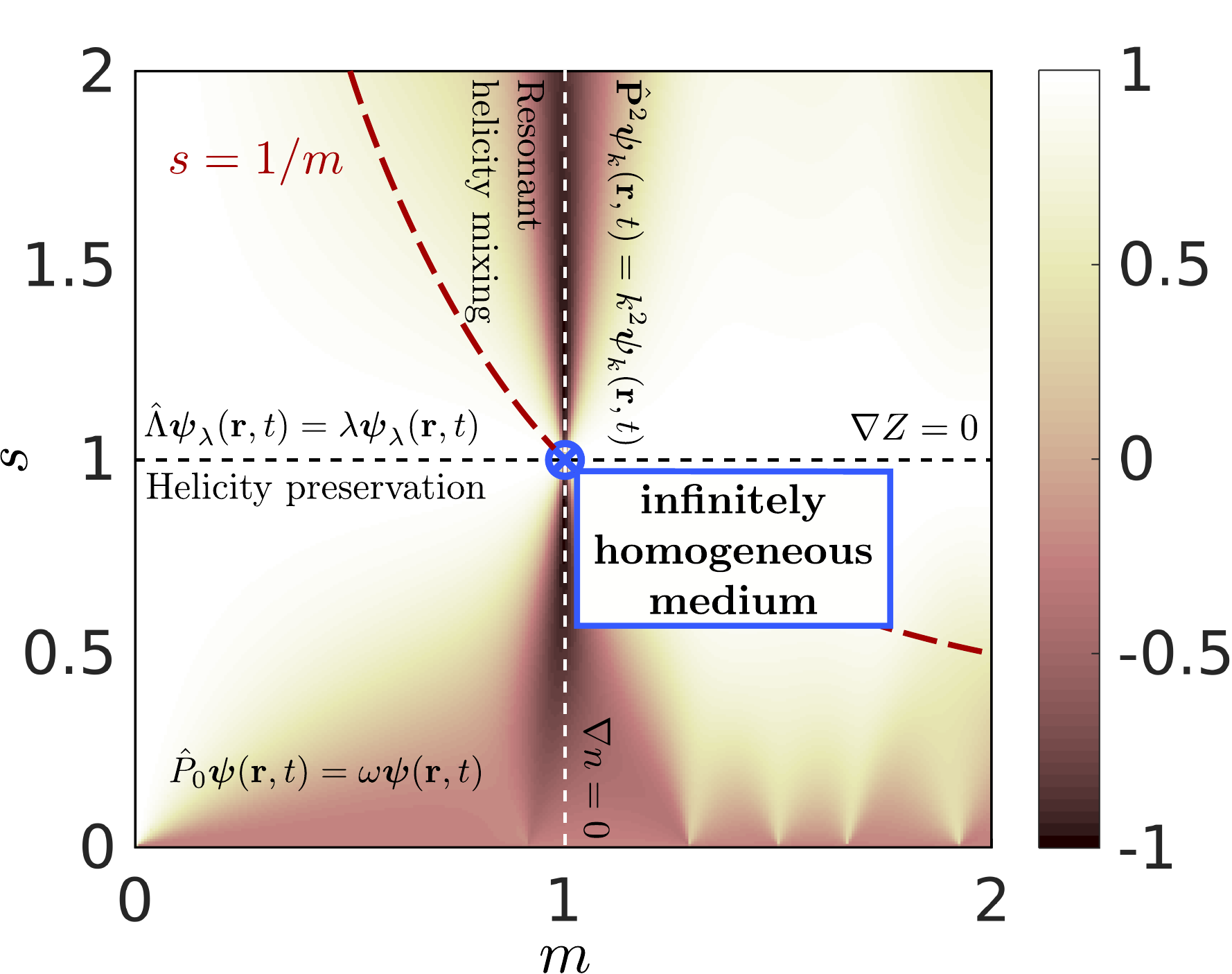}
    \caption{Emergence of the Kerker phenomena in a magnetic sphere embedded in a homogeneous surrounding. The helicity expectation value, $\langle\Lambda\rangle$, quantifies the helicity of the scattered electromagnetic field. $s$ is the impedance contrast and $m$ the refractive index contrast. Frequency, $\omega$, is preserved at every point due to the static nature of the piecewise system. Helicity, is preserved whenever the impedance is constant (horizontal dashed line). Square of linear momentum, is preserved whenever the refractive index is constant (vertical dashed line). The singular point $[m, s] = [1, 1]$ represents an infinitely homogeneous medium.}
    \label{FigE3}
\end{figure}

In Fig. \ref{FigE3}, we show the analysis of conserved quantities associated with the helicity map of a magnetic sphere with size parameter $x = 3$. Let us briefly remind that $\langle \Lambda \rangle$ represents the helicity expectation value, which is an observable that determines whether electromagnetic helicity of the incident wave is conserved, $\langle\Lambda\rangle = 1$, or completely flipped, $\langle\Lambda\rangle = -1$, upon scattering \cite{Correlations}. Along the vertical axis we tune the impedance contrast, $s$, which is the ratio between the impedances of the sphere and the surrounding medium. In the horizontal axis, we fix the refractive index contrast, $m$, which is the ratio between the refractive indices. Due to the static nature of the media involved, all the solutions represented in the figure are constructed as eigenstates of the generator of time-translations, $\hat{P}_0$. Then, under the impedance matching condition, along the line $s = 1$, helicity is conserved and, thus, solutions are eigenstates of $\hat{\Lambda}$. In addition to this, under the index matching condition, whenever $m = 1$, square of linear momentum is conserved and, as a result, solutions are eigenstates of $\hat{\mathbf{P}}^2$ operator. Note that, whenever $m = 1$, the helicity of the scattered field is almost completely flipped (see Fig. \ref{FigE3}) and, thus, this condition has been denoted as the resonant helicity mixing condition \cite{Resonant}. On the other hand, the line $s = 1/m$ represents the response of nonmagnetic materials, i.e. those for which $\mu = 1$. Finally, in the singular point $[m, s] = [1,1]$ of the colormap solutions are eigenstates of all three operators $\hat{P}_0$, $\hat{\Lambda}$ and $\hat{\mathbf{P}}^2$. This is because the particular case $m = s = 1$, represents an infinitely homogeneous medium and, by means of Eqs. \eqref{CasimirH}-\eqref{CasimirP2}, electromagnetic waves are constructed as eigenstates of the three Casimir operators of $P_{\scriptscriptstyle{{3,1}}}$.

The Kerker phenomena can also be interpreted making use of Maxwell's equations as expressed in Eq. \eqref{Max_InRS1}-\eqref{Max_InRS2}. Indeed, under both matching conditions, i.e. $\nabla Z = 0$ and $\nabla n = 0$, the environment ceases to be inhomogeneous and it becomes infinitely homogeneous. In this situation, Maxwell's equations converge to the analytical form dictated by Eqs. \eqref{CasimirH}-\eqref{CasimirP2}. In other words, electromagnetic wave solutions are constructed as eigenstates of the operators $\hat{P}_0$, $\hat{\Lambda}$ and $\hat{\mathbf{P}}^2$. This situation is equivalent to the one represented by $[m,s] = [1, 1]$ point in Fig. \ref{FigE3}. And, then, it is clear that Eqs. \eqref{Max_InRS1}-\eqref{Max_InRS2} indicate two preferential directions in which the homogeneity of space maybe broken. These two preferential directions are associated with the two matching conditions. On the one hand, homogeneity of space may be broken while still keeping the impedance constant. In piecewise homogeneous media, this leads to the conservation of both $\hat{P}_0$ and $\hat{\Lambda}$ and the situation is equivalent to the $s = 1$ line in Fig. \ref{FigE3}. On the other hand, homogeneity of space may also be broken while keeping the refractive index constant. In piecewise homogeneous media, this leads to the conservation of $\hat{P}_0$ and $\hat{\mathbf{P}}^2$ and the situation is equivalent to the $m = 1$ line in Fig. \ref{FigE3}.

In our view, this analysis provides a fundamental interpretation of Maxwell's equations in inhomogeneous media as expressed in Eqs. \eqref{Max_InRS1}-\eqref{Max_InRS2}. Please note that such a particular form of expressing the equations is obtained through a simple change of basis. Indeed, instead of the usual electric, $\mathbf{E}(\mathbf{r}, t)$, and magnetic, $\mathbf{H}(\mathbf{r}, t)$, fields we rewrite Maxwell's equations in terms of the monochromatic RS vector, $\boldsymbol{\Phi}^\lambda(\mathbf{r}, t)$, which is linked with the UIRs of the $P_{\scriptscriptstyle{{3,1}}}$ subgroup of the Poincaré group (see Section III). As a result, we obtain a completely equivalent form of the equations in which instead of the usual material parameters, derivatives of the local impedance, $Z(\mathbf{r})$, and refractive index, $n(\mathbf{r})$, appear. Finally, we find that these material parameters are closely related with the Casimir invariants of $P_{\scriptscriptstyle{{3,1}}}$. It really seems like the connection of Maxwell's equations with space-time symmetries is clarified when expressing them in terms of the RS vector. In this line, Kerker phenomena represent an expression of such a fundamental connection in a particular electromagnetic scattering problem.

\section{Conclusion} \label{sec:6}

In conclusion, we have revisited the link of Maxwell's equations with the theory of representations of continuous groups. Starting from the Poincaré group, we have shown that its subgroups may be employed to analyze conserved quantities and wave solutions in different material environments. It is the first time that this procedure, which had previously been employed in the framework of quantum mechanics, is systematically applied to the study of electromagnetic waves.

In this context, we have obtained a set of fundamental results. First, we have shown that Bialynicki-Birula's photon wave function in vacuum, $\boldsymbol{\Psi}^\lambda(\mathbf{r}, t)$, can be constructed from pure group theoretical arguments. Second, we have shown that Maxwell's equations in infinitely homogeneous media can be derived from the Casimir invariants of $P_{\scriptscriptstyle{{3,1}}}$ subgroup of the Poincaré group. As a result, we have identified the fundamental role that the monochromatic RS vector, $\boldsymbol{\Phi}^\lambda(\mathbf{r}, t)$, plays in the description of electromagnetic waves propagating in matter. Third, we have seen that the matching conditions in piecewise media are associated with the conservation of two Casimir invariants of $P_{\scriptscriptstyle{{3,1}}}$, i.e. helicity and square of linear momentum. Finally, we have shown that Kerker phenomena constitute a particular example of how the matching conditions emerge in piecewise homogeneous media.

Our contribution not only gives a sound mathematical foundation to the origin of the Kerker phenomena, but also invites to the search of new effects based on similar group theoretical arguments. In addition, we have identified the monochromatic RS vector in Eq. \eqref{PWF_medium} as the fundamental object associated with the description of electromagnetic waves propagating in matter. This may be useful to obtain more insightful descriptions of optical phenomena in piecewise homogeneous systems.

\section*{Acknowledgments}
J.L.A. acknowledges discussions with Jorge Olmos-Trigo. J.L.A., C.D. and A.G.E acknowledge support from Project No. PID2019-109905GA-C22 of the Spanish Ministerio de Ciencia, Innovaci\'on y Universidades (MICIU), IKUR Strategy under the collaboration agreement between Ikerbasque Foundation and DIPC on behalf of the Department of Education of the Basque Government, Gipuzkoa Quantum: QUAN-000021-01 project from Diputacion Foral de Gipuzkoa, Programa de ayudas de apoyo a los agentes de la Red Vasca de Ciencia, Tecnolog\'ia e Innovaci\'on acreditados en la categor\'ia de Centros de Investigaci\'on B\'asica y de Excelencia (Programa BERC) from the Departamento de Universidades e Investigaci\'on del Gobierno Vasco and Centros Severo Ochoa AEI/CEX2018-000867-S from the Spanish Ministerio de Ciencia e Innovaci\'on. G.M.T received funding from the IKUR Strategy under the collaboration agreement between Ikerbasque Foundation and DIPC/MPC on behalf of the Department of Education of the Basque Government.

\appendix
\section{Boosts acting over a monochromatic plane wave with well-defined helicity} \label{AppendixA}

Let us analyze the action of the Lorentz boost in the $OZ$ direction, $\hat{L}_z(\xi)$, over the standard vector. In general, if a boost is carried out in the $\mathbf{b}/|\mathbf{b}|$ direction, a generic electromagnetic field with well-defined helicity, $\mathbf{G}_\pm$, is transformed as \cite{Jackson, IvanThesis}:
\begin{equation}
    \label{BoostEq}
    \mathbf{G}'_\pm = \gamma\left( \mathbf{G}_\pm \mp i\mathbf{b}\times\mathbf{G}_\pm \right) - \frac{\gamma^2}{\gamma + 1}\mathbf{b}\left( \mathbf{b}\cdot\mathbf{G}_\pm \right),
\end{equation}
where $|\mathbf{b}| = v$ and $\gamma = (1 - |\mathbf{b}|^2)^{-1/2}$. Here, we are interested the transformation properties of the field $\mathbf{G}_\pm = \mathbf{u}_\pm e^{i|\mathbf{k}_l|(z - t)}$, with $\mathbf{u}_\pm = (1, \pm i, 0)$. Moreover, as the boost is along the $OZ$ axis, we need to fix $\mathbf{b}/|\mathbf{b}| = \hat{z}$ and, as a result, we have that $i\mathbf{b}\times\mathbf{u}_\pm = \pm|\mathbf{b}|\mathbf{u}_\pm$. It can also be checked that $\mathbf{b}$ and $\mathbf{G}_\pm$ are orthogonal, i.e. $\mathbf{b}\cdot\mathbf{G}_\pm = 0$, which makes the second term in Eq. \eqref{BoostEq} vanish. Finally, applying the inverse Lorentz transformation to the coordinates, i.e. from $(\mathbf{r}, t)$ to $(\mathbf{r}', t')$, the standard vector finally results in:
\begin{equation}
    \mathbf{G}'_\pm = \gamma(1 - |\mathbf{b}|)\mathbf{u}_\pm e^{i\gamma(1 - |\mathbf{b}|)|\mathbf{k}_l|(z' - t')}.
\end{equation}

The expression in the equation above shows that apart from changing its amplitude, the standard vector also changes the modulus of its linear momentum vector as:
\begin{equation}
    |\mathbf{k}'| = \gamma(1 - |\mathbf{b}|)|\mathbf{k}_l| = (\cosh\xi - \sinh\xi)|\mathbf{k}_l| = e^{-\xi}|\mathbf{k}_l|.
    \label{kchange}
\end{equation}
Please note that different sign conventions are considered in Wu-Ki Tung and Jackson for the Lorentz boosts (see, for instance, Eq. 10.1-9 in Wu-Ki Tung and Eq. 11.21 in Jackson). The result in Eq. \eqref{kchange} implies that the boost parameter $\xi$ controls the modulus of the linear momentum, $\mathbf{k}'$, of the monochromatic plane wave for a fixed $|\mathbf{k}_l|$. As the frequency and the modulus of linear momentum are proportional, one finally gets that $\omega$ is also modulated by the parameter $\xi$. Importantly, note that the boost $\hat{L}_z(\xi)$ does not modify either the helicity or the direction of propagation of the standard vector.

\section{Conservation of $\hat{\mathbf{P}}^2$ for quantum non-relativistic massive particles} \label{AppendixB}
In this appendix, we show that the conservation of $\hat{\mathbf{P}}^2$ for quantum non-relativistic massive particles does not respond to the restoration of a symmetry, but just to the properties of particular scattering potentials.

In our previous discussion we have been focused on the description of electromagnetic waves in different media. However, the mathematical tool that has been employed, i.e. the symmetry breaking principle applied to the Poincaré group, does not mind about the nature of the particle it is describing. In other words, the mathematical principle operates regardless of whether the physical theory represents a massless or massive particle \cite{SymmetryBreak2}. In this line, we have shown that the identities given by Eqs. \eqref{CasimirH}-\eqref{CasimirP2} can also represents the dynamic equations of a non-relativistic massive particle, i.e. the time-independent Schrödinger's equation with a constant potential. This implies that the UIRs of $P_{\scriptscriptstyle{{3,1}}}$ and its subgroups also play a fundamental role in the determination of the wave functions of massive particles \cite{SymmetryBreak3}. In the upcoming, we focus on the role of the square of linear momentum operator: we seek a physical situation for massive particles in which $\hat{\mathbf{P}}^2$ is a conserved quantity, regardless of the geometry of the problem. 

Such a situation would be reached by the scattering of a massive particle in an inhomogeneous potential, $\hat{V}(\mathbf{r})$, for which the solutions still remain being eigenstates of the $\hat{\mathbf{P}}^2$ operator, regardless of the symmetries of the problem. This would similarly imply that a Casimir operator of $P_{\scriptscriptstyle{3,1}}$ group can also be preserved in inhomogeneous environments. Interestingly, there is a potential that fulfills the requirements above: the generalized quantum hard sphere potential, which can be defined as:
\begin{equation}
    \hat{V}_{hs}(\mathbf{r}) = \left\{ \begin{array}{lcc}
             \infty &   \text{if}  & \mathbf{r} \in U_i \\
             \\ V_0 &  \text{else},
             \end{array}
   \right.
\end{equation}
where $U_i$ with $i=1, 2,...,N$, are $N$ arbitrary regions of the three dimensional space and $V_0$ represents a finite and constant potential energy. Crucially, due to the arbitrariness of the $U_i$ regions, such a potential can be constructed with any possible geometry. The solutions of the time-independent Schrödinger's equations in such a potential, $\psi_{hs}(\mathbf{r})$, are most generally determined by the following constrains
\begin{equation}
    \left\{ \begin{array}{lcc}
             \psi_{hs}(\mathbf{r}, t) = 0 &   \text{if}  & \mathbf{r} \in U_i \\
             \\ \hat{\mathbf{P}}^2 \psi_{hs}(\mathbf{r}, t) = 2m(E-V_0) \psi_{hs}(\mathbf{r}, t) &  \text{else}.
             \end{array}
    \right.
\end{equation}
As a result, it can be checked that the wave functions which are solutions to the generalized quantum hard sphere potential fulfill: $\hat{\mathbf{P}}^2\psi_{hs}(\mathbf{r}, t) = p^2\psi_{hs}(\mathbf{r}, t)$, $\forall~\mathbf{r}, t$ with $p = \sqrt{2m(E-V_0)}$.

From the properties above, it directly follows that the square of linear momentum is a conserved magnitude in the scattering with a hard sphere potential. Indeed, one can show that the expected value of $\hat{\mathbf{P}}^2$ remains constant for such a potential, i.e.
\begin{equation}
    \frac{d}{dt} \left(\int d\mathbf{r}~\psi^*_{hs}(\mathbf{r}, t)\left[\hat{\mathbf{P}}^2 \psi_{hs}(\mathbf{r}, t)\right]\right) = 0.
\end{equation}
This implies that if at an initial instant solutions are eigenstates of $\hat{\mathbf{P}}^2$, this property continues to hold in the curse of time. In other words, one usually says that $p$ is a good quantum number. Note that exactly the same holds for the kinetic energy, $\hat{T} = \hat{\mathbf{P}}^2/2m$, as the mass is a fixed parameter in non-relativistic quantum mechanics. Both in quantum and classical scattering theory the conservation of kinetic energy is just invoked as a property of certain type of interactions. As a result, we conclude that the preservation of the square of linear momentum is not, in general, associated with the restoration of a space-time symmetry.

\bibliography{mybib}

\begin{thebibliography}{59}%
\makeatletter
\providecommand \@ifxundefined [1]{%
 \@ifx{#1\undefined}
}%
\providecommand \@ifnum [1]{%
 \ifnum #1\expandafter \@firstoftwo
 \else \expandafter \@secondoftwo
 \fi
}%
\providecommand \@ifx [1]{%
 \ifx #1\expandafter \@firstoftwo
 \else \expandafter \@secondoftwo
 \fi
}%
\providecommand \natexlab [1]{#1}%
\providecommand \enquote  [1]{``#1''}%
\providecommand \bibnamefont  [1]{#1}%
\providecommand \bibfnamefont [1]{#1}%
\providecommand \citenamefont [1]{#1}%
\providecommand \href@noop [0]{\@secondoftwo}%
\providecommand \href [0]{\begingroup \@sanitize@url \@href}%
\providecommand \@href[1]{\@@startlink{#1}\@@href}%
\providecommand \@@href[1]{\endgroup#1\@@endlink}%
\providecommand \@sanitize@url [0]{\catcode `\\12\catcode `\$12\catcode
  `\&12\catcode `\#12\catcode `\^12\catcode `\_12\catcode `\%12\relax}%
\providecommand \@@startlink[1]{}%
\providecommand \@@endlink[0]{}%
\providecommand \url  [0]{\begingroup\@sanitize@url \@url }%
\providecommand \@url [1]{\endgroup\@href {#1}{\urlprefix }}%
\providecommand \urlprefix  [0]{URL }%
\providecommand \Eprint [0]{\href }%
\providecommand \doibase [0]{http://dx.doi.org/}%
\providecommand \selectlanguage [0]{\@gobble}%
\providecommand \bibinfo  [0]{\@secondoftwo}%
\providecommand \bibfield  [0]{\@secondoftwo}%
\providecommand \translation [1]{[#1]}%
\providecommand \BibitemOpen [0]{}%
\providecommand \bibitemStop [0]{}%
\providecommand \bibitemNoStop [0]{.\EOS\space}%
\providecommand \EOS [0]{\spacefactor3000\relax}%
\providecommand \BibitemShut  [1]{\csname bibitem#1\endcsname}%
\let\auto@bib@innerbib\@empty
\bibitem [{\citenamefont {Kerker}\ \emph {et~al.}(1983)\citenamefont {Kerker},
  \citenamefont {Wang},\ and\ \citenamefont {Giles}}]{Kerker}%
  \BibitemOpen
  \bibfield  {author} {\bibinfo {author} {\bibfnamefont {M.}~\bibnamefont
  {Kerker}}, \bibinfo {author} {\bibfnamefont {D.-S.}\ \bibnamefont {Wang}}, \
  and\ \bibinfo {author} {\bibfnamefont {C.~L.}\ \bibnamefont {Giles}},\
  }\href@noop {} {\bibfield  {journal} {\bibinfo  {journal} {J. Opt. Soc. Am.}\
  }\textbf {\bibinfo {volume} {73}},\ \bibinfo {pages} {765} (\bibinfo {year}
  {1983})}\BibitemShut {NoStop}%
\bibitem [{\citenamefont {Giles}\ and\ \citenamefont {Wild}(1982)}]{GilesWild}%
  \BibitemOpen
  \bibfield  {author} {\bibinfo {author} {\bibfnamefont {C.~L.}\ \bibnamefont
  {Giles}}\ and\ \bibinfo {author} {\bibfnamefont {W.~J.}\ \bibnamefont
  {Wild}},\ }\href@noop {} {\bibfield  {journal} {\bibinfo  {journal} {Applied
  Physics Letters}\ }\textbf {\bibinfo {volume} {40}},\ \bibinfo {pages} {210}
  (\bibinfo {year} {1982})}\BibitemShut {NoStop}%
\bibitem [{\citenamefont {Nieto-Vesperinas}\ \emph
  {et~al.}(2011{\natexlab{a}})\citenamefont {Nieto-Vesperinas}, \citenamefont
  {G{\'o}mez-Medina},\ and\ \citenamefont {S{\'a}enz}}]{KerkerApp0}%
  \BibitemOpen
  \bibfield  {author} {\bibinfo {author} {\bibfnamefont {M.}~\bibnamefont
  {Nieto-Vesperinas}}, \bibinfo {author} {\bibfnamefont {R.}~\bibnamefont
  {G{\'o}mez-Medina}}, \ and\ \bibinfo {author} {\bibfnamefont {J.~J.}\
  \bibnamefont {S{\'a}enz}},\ }\href@noop {} {\bibfield  {journal} {\bibinfo
  {journal} {J. Opt. Soc. Am. A}\ }\textbf {\bibinfo {volume} {28}},\ \bibinfo
  {pages} {54} (\bibinfo {year} {2011}{\natexlab{a}})}\BibitemShut {NoStop}%
\bibitem [{\citenamefont {Garc\'{\i}a-Etxarri}\ and\ \citenamefont
  {Dionne}(2013)}]{KerkerApp1}%
  \BibitemOpen
  \bibfield  {author} {\bibinfo {author} {\bibfnamefont {A.}~\bibnamefont
  {Garc\'{\i}a-Etxarri}}\ and\ \bibinfo {author} {\bibfnamefont {J.~A.}\
  \bibnamefont {Dionne}},\ }\href@noop {} {\bibfield  {journal} {\bibinfo
  {journal} {Phys. Rev. B}\ }\textbf {\bibinfo {volume} {87}},\ \bibinfo
  {pages} {235409} (\bibinfo {year} {2013})}\BibitemShut {NoStop}%
\bibitem [{\citenamefont {Pfeiffer}\ and\ \citenamefont
  {Grbic}(2013)}]{KerkerApp2}%
  \BibitemOpen
  \bibfield  {author} {\bibinfo {author} {\bibfnamefont {C.}~\bibnamefont
  {Pfeiffer}}\ and\ \bibinfo {author} {\bibfnamefont {A.}~\bibnamefont
  {Grbic}},\ }\href@noop {} {\bibfield  {journal} {\bibinfo  {journal} {Phys.
  Rev. Lett.}\ }\textbf {\bibinfo {volume} {110}},\ \bibinfo {pages} {197401}
  (\bibinfo {year} {2013})}\BibitemShut {NoStop}%
\bibitem [{\citenamefont {Person}\ \emph {et~al.}(2013)\citenamefont {Person},
  \citenamefont {Jain}, \citenamefont {Lapin}, \citenamefont {S{\'a}enz},
  \citenamefont {Wicks},\ and\ \citenamefont {Novotny}}]{KerkerRef01}%
  \BibitemOpen
  \bibfield  {author} {\bibinfo {author} {\bibfnamefont {S.}~\bibnamefont
  {Person}}, \bibinfo {author} {\bibfnamefont {M.}~\bibnamefont {Jain}},
  \bibinfo {author} {\bibfnamefont {Z.}~\bibnamefont {Lapin}}, \bibinfo
  {author} {\bibfnamefont {J.~J.}\ \bibnamefont {S{\'a}enz}}, \bibinfo {author}
  {\bibfnamefont {G.}~\bibnamefont {Wicks}}, \ and\ \bibinfo {author}
  {\bibfnamefont {L.}~\bibnamefont {Novotny}},\ }\href@noop {} {\bibfield
  {journal} {\bibinfo  {journal} {Nano Letters}\ }\textbf {\bibinfo {volume}
  {13}},\ \bibinfo {pages} {1806} (\bibinfo {year} {2013})}\BibitemShut
  {NoStop}%
\bibitem [{\citenamefont {Fu}\ \emph {et~al.}(2013)\citenamefont {Fu},
  \citenamefont {Kuznetsov}, \citenamefont {Miroshnichenko}, \citenamefont
  {Yu},\ and\ \citenamefont {Luk’yanchuk}}]{KerkerRef1}%
  \BibitemOpen
  \bibfield  {author} {\bibinfo {author} {\bibfnamefont {Y.~H.}\ \bibnamefont
  {Fu}}, \bibinfo {author} {\bibfnamefont {A.~I.}\ \bibnamefont {Kuznetsov}},
  \bibinfo {author} {\bibfnamefont {A.~E.}\ \bibnamefont {Miroshnichenko}},
  \bibinfo {author} {\bibfnamefont {Y.~F.}\ \bibnamefont {Yu}}, \ and\ \bibinfo
  {author} {\bibfnamefont {B.}~\bibnamefont {Luk’yanchuk}},\ }\href@noop {}
  {\bibfield  {journal} {\bibinfo  {journal} {Nature Communications}\ }\textbf
  {\bibinfo {volume} {4}} (\bibinfo {year} {2013})}\BibitemShut {NoStop}%
\bibitem [{\citenamefont {Alaee}\ \emph {et~al.}(2015)\citenamefont {Alaee},
  \citenamefont {Filter}, \citenamefont {Lehr}, \citenamefont {Lederer},\ and\
  \citenamefont {Rockstuhl}}]{Kerker11}%
  \BibitemOpen
  \bibfield  {author} {\bibinfo {author} {\bibfnamefont {R.}~\bibnamefont
  {Alaee}}, \bibinfo {author} {\bibfnamefont {R.}~\bibnamefont {Filter}},
  \bibinfo {author} {\bibfnamefont {D.}~\bibnamefont {Lehr}}, \bibinfo {author}
  {\bibfnamefont {F.}~\bibnamefont {Lederer}}, \ and\ \bibinfo {author}
  {\bibfnamefont {C.}~\bibnamefont {Rockstuhl}},\ }\href@noop {} {\bibfield
  {journal} {\bibinfo  {journal} {Opt. Lett.}\ }\textbf {\bibinfo {volume}
  {40}},\ \bibinfo {pages} {2645} (\bibinfo {year} {2015})}\BibitemShut
  {NoStop}%
\bibitem [{\citenamefont {Babicheva}\ and\ \citenamefont
  {Evlyukhin}(2017)}]{KerkerApp4}%
  \BibitemOpen
  \bibfield  {author} {\bibinfo {author} {\bibfnamefont {V.~E.}\ \bibnamefont
  {Babicheva}}\ and\ \bibinfo {author} {\bibfnamefont {A.~B.}\ \bibnamefont
  {Evlyukhin}},\ }\href@noop {} {\bibfield  {journal} {\bibinfo  {journal}
  {Laser \& Photonics Reviews}\ }\textbf {\bibinfo {volume} {11}},\ \bibinfo
  {pages} {1700132} (\bibinfo {year} {2017})}\BibitemShut {NoStop}%
\bibitem [{\citenamefont {Bag}\ \emph {et~al.}(2018)\citenamefont {Bag},
  \citenamefont {Neugebauer}, \citenamefont {Wo\ifmmode~\acute{z}\else
  \'{z}\fi{}niak}, \citenamefont {Leuchs},\ and\ \citenamefont
  {Banzer}}]{Kerker13}%
  \BibitemOpen
  \bibfield  {author} {\bibinfo {author} {\bibfnamefont {A.}~\bibnamefont
  {Bag}}, \bibinfo {author} {\bibfnamefont {M.}~\bibnamefont {Neugebauer}},
  \bibinfo {author} {\bibfnamefont {P.}~\bibnamefont {Wo\ifmmode~\acute{z}\else
  \'{z}\fi{}niak}}, \bibinfo {author} {\bibfnamefont {G.}~\bibnamefont
  {Leuchs}}, \ and\ \bibinfo {author} {\bibfnamefont {P.}~\bibnamefont
  {Banzer}},\ }\href@noop {} {\bibfield  {journal} {\bibinfo  {journal} {Phys.
  Rev. Lett.}\ }\textbf {\bibinfo {volume} {121}},\ \bibinfo {pages} {193902}
  (\bibinfo {year} {2018})}\BibitemShut {NoStop}%
\bibitem [{\citenamefont {Barhom}\ \emph {et~al.}(2019)\citenamefont {Barhom},
  \citenamefont {Machnev}, \citenamefont {Noskov}, \citenamefont {Goncharenko},
  \citenamefont {Gurvitz}, \citenamefont {Timin}, \citenamefont {Shkoldin},
  \citenamefont {Koniakhin}, \citenamefont {Koval}, \citenamefont {Zyuzin},
  \citenamefont {Shalin}, \citenamefont {Shishkin},\ and\ \citenamefont
  {Ginzburg}}]{Kerker14}%
  \BibitemOpen
  \bibfield  {author} {\bibinfo {author} {\bibfnamefont {H.}~\bibnamefont
  {Barhom}}, \bibinfo {author} {\bibfnamefont {A.~A.}\ \bibnamefont {Machnev}},
  \bibinfo {author} {\bibfnamefont {R.~E.}\ \bibnamefont {Noskov}}, \bibinfo
  {author} {\bibfnamefont {A.}~\bibnamefont {Goncharenko}}, \bibinfo {author}
  {\bibfnamefont {E.~A.}\ \bibnamefont {Gurvitz}}, \bibinfo {author}
  {\bibfnamefont {A.~S.}\ \bibnamefont {Timin}}, \bibinfo {author}
  {\bibfnamefont {V.~A.}\ \bibnamefont {Shkoldin}}, \bibinfo {author}
  {\bibfnamefont {S.~V.}\ \bibnamefont {Koniakhin}}, \bibinfo {author}
  {\bibfnamefont {O.~Y.}\ \bibnamefont {Koval}}, \bibinfo {author}
  {\bibfnamefont {M.~V.}\ \bibnamefont {Zyuzin}}, \bibinfo {author}
  {\bibfnamefont {A.~S.}\ \bibnamefont {Shalin}}, \bibinfo {author}
  {\bibfnamefont {I.~I.}\ \bibnamefont {Shishkin}}, \ and\ \bibinfo {author}
  {\bibfnamefont {P.}~\bibnamefont {Ginzburg}},\ }\href@noop {} {\bibfield
  {journal} {\bibinfo  {journal} {Nano Letters}\ }\textbf {\bibinfo {volume}
  {19}},\ \bibinfo {pages} {7062} (\bibinfo {year} {2019})}\BibitemShut
  {NoStop}%
\bibitem [{\citenamefont {Lasa-Alonso}\ \emph
  {et~al.}(2020{\natexlab{a}})\citenamefont {Lasa-Alonso}, \citenamefont
  {Abujetas}, \citenamefont {Nodar}, \citenamefont {Dionne}, \citenamefont
  {S\'aenz}, \citenamefont {Molina-Terriza}, \citenamefont {Aizpurua},\ and\
  \citenamefont {Garc\'ia-Etxarri}}]{ACSLasa}%
  \BibitemOpen
  \bibfield  {author} {\bibinfo {author} {\bibfnamefont {J.}~\bibnamefont
  {Lasa-Alonso}}, \bibinfo {author} {\bibfnamefont {D.~R.}\ \bibnamefont
  {Abujetas}}, \bibinfo {author} {\bibfnamefont {A.}~\bibnamefont {Nodar}},
  \bibinfo {author} {\bibfnamefont {J.~A.}\ \bibnamefont {Dionne}}, \bibinfo
  {author} {\bibfnamefont {J.~J.}\ \bibnamefont {S\'aenz}}, \bibinfo {author}
  {\bibfnamefont {G.}~\bibnamefont {Molina-Terriza}}, \bibinfo {author}
  {\bibfnamefont {J.}~\bibnamefont {Aizpurua}}, \ and\ \bibinfo {author}
  {\bibfnamefont {A.}~\bibnamefont {Garc\'ia-Etxarri}},\ }\href@noop {}
  {\bibfield  {journal} {\bibinfo  {journal} {ACS Photonics}\ }\textbf
  {\bibinfo {volume} {7}},\ \bibinfo {pages} {2978–2986} (\bibinfo {year}
  {2020}{\natexlab{a}})}\BibitemShut {NoStop}%
\bibitem [{\citenamefont {Xu}\ \emph {et~al.}(2020)\citenamefont {Xu},
  \citenamefont {Nieto-Vesperinas}, \citenamefont {Qiu}, \citenamefont {Liu},
  \citenamefont {Gao}, \citenamefont {Zhang},\ and\ \citenamefont
  {Li}}]{Kerker18}%
  \BibitemOpen
  \bibfield  {author} {\bibinfo {author} {\bibfnamefont {X.}~\bibnamefont
  {Xu}}, \bibinfo {author} {\bibfnamefont {M.}~\bibnamefont
  {Nieto-Vesperinas}}, \bibinfo {author} {\bibfnamefont {C.-W.}\ \bibnamefont
  {Qiu}}, \bibinfo {author} {\bibfnamefont {X.}~\bibnamefont {Liu}}, \bibinfo
  {author} {\bibfnamefont {D.}~\bibnamefont {Gao}}, \bibinfo {author}
  {\bibfnamefont {Y.}~\bibnamefont {Zhang}}, \ and\ \bibinfo {author}
  {\bibfnamefont {B.}~\bibnamefont {Li}},\ }\href@noop {} {\bibfield  {journal}
  {\bibinfo  {journal} {Laser \& Photonics Reviews}\ }\textbf {\bibinfo
  {volume} {14}},\ \bibinfo {pages} {1900265} (\bibinfo {year}
  {2020})}\BibitemShut {NoStop}%
\bibitem [{\citenamefont {Lasa-Alonso}\ \emph
  {et~al.}(2020{\natexlab{b}})\citenamefont {Lasa-Alonso}, \citenamefont
  {Molezuelas-Ferreras}, \citenamefont {Varga}, \citenamefont
  {Garc{\'{\i}}a-Etxarri}, \citenamefont {Giedke},\ and\ \citenamefont
  {Molina-Terriza}}]{SymProt}%
  \BibitemOpen
  \bibfield  {author} {\bibinfo {author} {\bibfnamefont {J.}~\bibnamefont
  {Lasa-Alonso}}, \bibinfo {author} {\bibfnamefont {M.}~\bibnamefont
  {Molezuelas-Ferreras}}, \bibinfo {author} {\bibfnamefont {J.~J.~M.}\
  \bibnamefont {Varga}}, \bibinfo {author} {\bibfnamefont {A.}~\bibnamefont
  {Garc{\'{\i}}a-Etxarri}}, \bibinfo {author} {\bibfnamefont {G.}~\bibnamefont
  {Giedke}}, \ and\ \bibinfo {author} {\bibfnamefont {G.}~\bibnamefont
  {Molina-Terriza}},\ }\href@noop {} {\bibfield  {journal} {\bibinfo  {journal}
  {New Journal of Physics}\ }\textbf {\bibinfo {volume} {22}},\ \bibinfo
  {pages} {123010} (\bibinfo {year} {2020}{\natexlab{b}})}\BibitemShut
  {NoStop}%
\bibitem [{\citenamefont {Olmos-Trigo}\ \emph
  {et~al.}(2020{\natexlab{a}})\citenamefont {Olmos-Trigo}, \citenamefont
  {Sanz-Fern\'andez}, \citenamefont {Abujetas}, \citenamefont {Lasa-Alonso},
  \citenamefont {de~Sousa}, \citenamefont {Garc\'ia-Etxarri}, \citenamefont
  {S\'anchez-Gil}, \citenamefont {Molina-Terriza},\ and\ \citenamefont
  {S\'aenz}}]{PRLJorge}%
  \BibitemOpen
  \bibfield  {author} {\bibinfo {author} {\bibfnamefont {J.}~\bibnamefont
  {Olmos-Trigo}}, \bibinfo {author} {\bibfnamefont {C.}~\bibnamefont
  {Sanz-Fern\'andez}}, \bibinfo {author} {\bibfnamefont {D.~R.}\ \bibnamefont
  {Abujetas}}, \bibinfo {author} {\bibfnamefont {J.}~\bibnamefont
  {Lasa-Alonso}}, \bibinfo {author} {\bibfnamefont {N.}~\bibnamefont
  {de~Sousa}}, \bibinfo {author} {\bibfnamefont {A.}~\bibnamefont
  {Garc\'ia-Etxarri}}, \bibinfo {author} {\bibfnamefont {J.~A.}\ \bibnamefont
  {S\'anchez-Gil}}, \bibinfo {author} {\bibfnamefont {G.}~\bibnamefont
  {Molina-Terriza}}, \ and\ \bibinfo {author} {\bibfnamefont {J.~J.}\
  \bibnamefont {S\'aenz}},\ }\href@noop {} {\bibfield  {journal} {\bibinfo
  {journal} {Physical Review Letters}\ }\textbf {\bibinfo {volume} {125}},\
  \bibinfo {pages} {073205} (\bibinfo {year} {2020}{\natexlab{a}})}\BibitemShut
  {NoStop}%
\bibitem [{\citenamefont {Fan}\ \emph {et~al.}(2021)\citenamefont {Fan},
  \citenamefont {Shadrivov}, \citenamefont {Miroshnichenko},\ and\
  \citenamefont {Padilla}}]{Kerker15}%
  \BibitemOpen
  \bibfield  {author} {\bibinfo {author} {\bibfnamefont {K.}~\bibnamefont
  {Fan}}, \bibinfo {author} {\bibfnamefont {I.~V.}\ \bibnamefont {Shadrivov}},
  \bibinfo {author} {\bibfnamefont {A.~E.}\ \bibnamefont {Miroshnichenko}}, \
  and\ \bibinfo {author} {\bibfnamefont {W.~J.}\ \bibnamefont {Padilla}},\
  }\href@noop {} {\bibfield  {journal} {\bibinfo  {journal} {Opt. Express}\
  }\textbf {\bibinfo {volume} {29}},\ \bibinfo {pages} {10518} (\bibinfo {year}
  {2021})}\BibitemShut {NoStop}%
\bibitem [{\citenamefont {Gerasimov}\ \emph {et~al.}(2021)\citenamefont
  {Gerasimov}, \citenamefont {Ershov}, \citenamefont {Bikbaev}, \citenamefont
  {Rasskazov}, \citenamefont {Isaev}, \citenamefont {Semina}, \citenamefont
  {Kostyukov}, \citenamefont {Zakomirnyi}, \citenamefont {Polyutov},\ and\
  \citenamefont {Karpov}}]{Kerker17}%
  \BibitemOpen
  \bibfield  {author} {\bibinfo {author} {\bibfnamefont {V.~S.}\ \bibnamefont
  {Gerasimov}}, \bibinfo {author} {\bibfnamefont {A.~E.}\ \bibnamefont
  {Ershov}}, \bibinfo {author} {\bibfnamefont {R.~G.}\ \bibnamefont {Bikbaev}},
  \bibinfo {author} {\bibfnamefont {I.~L.}\ \bibnamefont {Rasskazov}}, \bibinfo
  {author} {\bibfnamefont {I.~L.}\ \bibnamefont {Isaev}}, \bibinfo {author}
  {\bibfnamefont {P.~N.}\ \bibnamefont {Semina}}, \bibinfo {author}
  {\bibfnamefont {A.~S.}\ \bibnamefont {Kostyukov}}, \bibinfo {author}
  {\bibfnamefont {V.~I.}\ \bibnamefont {Zakomirnyi}}, \bibinfo {author}
  {\bibfnamefont {S.~P.}\ \bibnamefont {Polyutov}}, \ and\ \bibinfo {author}
  {\bibfnamefont {S.~V.}\ \bibnamefont {Karpov}},\ }\href@noop {} {\bibfield
  {journal} {\bibinfo  {journal} {Phys. Rev. B}\ }\textbf {\bibinfo {volume}
  {103}},\ \bibinfo {pages} {035402} (\bibinfo {year} {2021})}\BibitemShut
  {NoStop}%
\bibitem [{\citenamefont {Qin}\ \emph {et~al.}(2022)\citenamefont {Qin},
  \citenamefont {Zhang}, \citenamefont {Zheng}, \citenamefont {Xu},
  \citenamefont {Fu}, \citenamefont {Wang},\ and\ \citenamefont
  {Qin}}]{Kerker12}%
  \BibitemOpen
  \bibfield  {author} {\bibinfo {author} {\bibfnamefont {F.}~\bibnamefont
  {Qin}}, \bibinfo {author} {\bibfnamefont {Z.}~\bibnamefont {Zhang}}, \bibinfo
  {author} {\bibfnamefont {K.}~\bibnamefont {Zheng}}, \bibinfo {author}
  {\bibfnamefont {Y.}~\bibnamefont {Xu}}, \bibinfo {author} {\bibfnamefont
  {S.}~\bibnamefont {Fu}}, \bibinfo {author} {\bibfnamefont {Y.}~\bibnamefont
  {Wang}}, \ and\ \bibinfo {author} {\bibfnamefont {Y.}~\bibnamefont {Qin}},\
  }\href@noop {} {\bibfield  {journal} {\bibinfo  {journal} {Phys. Rev. Lett.}\
  }\textbf {\bibinfo {volume} {128}},\ \bibinfo {pages} {193901} (\bibinfo
  {year} {2022})}\BibitemShut {NoStop}%
\bibitem [{\citenamefont {Xiong}\ \emph {et~al.}(2022)\citenamefont {Xiong},
  \citenamefont {Ding}, \citenamefont {Lu},\ and\ \citenamefont
  {Li}}]{Kerker16}%
  \BibitemOpen
  \bibfield  {author} {\bibinfo {author} {\bibfnamefont {L.}~\bibnamefont
  {Xiong}}, \bibinfo {author} {\bibfnamefont {H.}~\bibnamefont {Ding}},
  \bibinfo {author} {\bibfnamefont {Y.}~\bibnamefont {Lu}}, \ and\ \bibinfo
  {author} {\bibfnamefont {G.}~\bibnamefont {Li}},\ }\href@noop {} {\bibfield
  {journal} {\bibinfo  {journal} {Journal of Physics D: Applied Physics}\
  }\textbf {\bibinfo {volume} {55}},\ \bibinfo {pages} {185106} (\bibinfo
  {year} {2022})}\BibitemShut {NoStop}%
\bibitem [{\citenamefont {Calkin}(1965)}]{Calkin}%
  \BibitemOpen
  \bibfield  {author} {\bibinfo {author} {\bibfnamefont {M.~G.}\ \bibnamefont
  {Calkin}},\ }\href@noop {} {\bibfield  {journal} {\bibinfo  {journal}
  {American Journal of Physics}\ }\textbf {\bibinfo {volume} {33}},\ \bibinfo
  {pages} {958} (\bibinfo {year} {1965})}\BibitemShut {NoStop}%
\bibitem [{\citenamefont {Fernandez-Corbaton}\ \emph
  {et~al.}(2013)\citenamefont {Fernandez-Corbaton}, \citenamefont
  {Zambrana-Puyalto}, \citenamefont {Tischler}, \citenamefont {Vidal},
  \citenamefont {Juan},\ and\ \citenamefont {Molina-Terriza}}]{PRLMolina}%
  \BibitemOpen
  \bibfield  {author} {\bibinfo {author} {\bibfnamefont {I.}~\bibnamefont
  {Fernandez-Corbaton}}, \bibinfo {author} {\bibfnamefont {X.}~\bibnamefont
  {Zambrana-Puyalto}}, \bibinfo {author} {\bibfnamefont {N.}~\bibnamefont
  {Tischler}}, \bibinfo {author} {\bibfnamefont {X.}~\bibnamefont {Vidal}},
  \bibinfo {author} {\bibfnamefont {M.~L.}\ \bibnamefont {Juan}}, \ and\
  \bibinfo {author} {\bibfnamefont {G.}~\bibnamefont {Molina-Terriza}},\
  }\href@noop {} {\bibfield  {journal} {\bibinfo  {journal} {Physical Review
  Letters}\ }\textbf {\bibinfo {volume} {111}},\ \bibinfo {pages} {060401}
  (\bibinfo {year} {2013})}\BibitemShut {NoStop}%
\bibitem [{\citenamefont {Zambrana-Puyalto}\ \emph {et~al.}(2013)\citenamefont
  {Zambrana-Puyalto}, \citenamefont {Fernandez-Corbaton}, \citenamefont {Juan},
  \citenamefont {Vidal},\ and\ \citenamefont
  {Molina-Terriza}}]{ZambranaKerker}%
  \BibitemOpen
  \bibfield  {author} {\bibinfo {author} {\bibfnamefont {X.}~\bibnamefont
  {Zambrana-Puyalto}}, \bibinfo {author} {\bibfnamefont {I.}~\bibnamefont
  {Fernandez-Corbaton}}, \bibinfo {author} {\bibfnamefont {M.~L.}\ \bibnamefont
  {Juan}}, \bibinfo {author} {\bibfnamefont {X.}~\bibnamefont {Vidal}}, \ and\
  \bibinfo {author} {\bibfnamefont {G.}~\bibnamefont {Molina-Terriza}},\
  }\href@noop {} {\bibfield  {journal} {\bibinfo  {journal} {Opt. Lett.}\
  }\textbf {\bibinfo {volume} {38}},\ \bibinfo {pages} {1857} (\bibinfo {year}
  {2013})}\BibitemShut {NoStop}%
\bibitem [{\citenamefont {Fernandez-Corbaton}(2013)}]{ForBackCorbaton}%
  \BibitemOpen
  \bibfield  {author} {\bibinfo {author} {\bibfnamefont {I.}~\bibnamefont
  {Fernandez-Corbaton}},\ }\href@noop {} {\bibfield  {journal} {\bibinfo
  {journal} {Optics Express}\ }\textbf {\bibinfo {volume} {21}},\ \bibinfo
  {pages} {29885} (\bibinfo {year} {2013})}\BibitemShut {NoStop}%
\bibitem [{\citenamefont {Lasa-Alonso}\ \emph {et~al.}(2022)\citenamefont
  {Lasa-Alonso}, \citenamefont {Olmos-Trigo}, \citenamefont {García-Etxarri},\
  and\ \citenamefont {Molina-Terriza}}]{Correlations}%
  \BibitemOpen
  \bibfield  {author} {\bibinfo {author} {\bibfnamefont {J.}~\bibnamefont
  {Lasa-Alonso}}, \bibinfo {author} {\bibfnamefont {J.}~\bibnamefont
  {Olmos-Trigo}}, \bibinfo {author} {\bibfnamefont {A.}~\bibnamefont
  {García-Etxarri}}, \ and\ \bibinfo {author} {\bibfnamefont {G.}~\bibnamefont
  {Molina-Terriza}},\ }\href@noop {} {\bibfield  {journal} {\bibinfo  {journal}
  {Mater. Adv.}\ }\textbf {\bibinfo {volume} {3}},\ \bibinfo {pages} {4179}
  (\bibinfo {year} {2022})}\BibitemShut {NoStop}%
\bibitem [{\citenamefont {Nieto-Vesperinas}\ \emph
  {et~al.}(2011{\natexlab{b}})\citenamefont {Nieto-Vesperinas}, \citenamefont
  {Gomez-Medina},\ and\ \citenamefont {Saenz}}]{NietoVesperinasK2}%
  \BibitemOpen
  \bibfield  {author} {\bibinfo {author} {\bibfnamefont {M.}~\bibnamefont
  {Nieto-Vesperinas}}, \bibinfo {author} {\bibfnamefont {R.}~\bibnamefont
  {Gomez-Medina}}, \ and\ \bibinfo {author} {\bibfnamefont {J.~J.}\
  \bibnamefont {Saenz}},\ }\href@noop {} {\bibfield  {journal} {\bibinfo
  {journal} {J. Opt. Soc. Am. A}\ }\textbf {\bibinfo {volume} {28}},\ \bibinfo
  {pages} {54} (\bibinfo {year} {2011}{\natexlab{b}})}\BibitemShut {NoStop}%
\bibitem [{\citenamefont {Olmos-Trigo}\ \emph
  {et~al.}(2020{\natexlab{b}})\citenamefont {Olmos-Trigo}, \citenamefont
  {Abujetas}, \citenamefont {Sanz-Fern\'andez}, \citenamefont {S\'anchez-Gil},\
  and\ \citenamefont {S\'aenz}}]{PRROlmos}%
  \BibitemOpen
  \bibfield  {author} {\bibinfo {author} {\bibfnamefont {J.}~\bibnamefont
  {Olmos-Trigo}}, \bibinfo {author} {\bibfnamefont {D.~R.}\ \bibnamefont
  {Abujetas}}, \bibinfo {author} {\bibfnamefont {C.}~\bibnamefont
  {Sanz-Fern\'andez}}, \bibinfo {author} {\bibfnamefont {J.~A.}\ \bibnamefont
  {S\'anchez-Gil}}, \ and\ \bibinfo {author} {\bibfnamefont {J.~J.}\
  \bibnamefont {S\'aenz}},\ }\href@noop {} {\bibfield  {journal} {\bibinfo
  {journal} {Physical Review Research}\ }\textbf {\bibinfo {volume} {2}},\
  \bibinfo {pages} {013225} (\bibinfo {year} {2020}{\natexlab{b}})}\BibitemShut
  {NoStop}%
\bibitem [{\citenamefont {Lasa-Alonso}\ \emph {et~al.}(2023)\citenamefont
  {Lasa-Alonso}, \citenamefont {Olmos-Trigo}, \citenamefont {Devescovi},
  \citenamefont {Hern\'andez}, \citenamefont {Garc\'{\i}a-Etxarri},\ and\
  \citenamefont {Molina-Terriza}}]{Resonant}%
  \BibitemOpen
  \bibfield  {author} {\bibinfo {author} {\bibfnamefont {J.}~\bibnamefont
  {Lasa-Alonso}}, \bibinfo {author} {\bibfnamefont {J.}~\bibnamefont
  {Olmos-Trigo}}, \bibinfo {author} {\bibfnamefont {C.}~\bibnamefont
  {Devescovi}}, \bibinfo {author} {\bibfnamefont {P.}~\bibnamefont
  {Hern\'andez}}, \bibinfo {author} {\bibfnamefont {A.}~\bibnamefont
  {Garc\'{\i}a-Etxarri}}, \ and\ \bibinfo {author} {\bibfnamefont
  {G.}~\bibnamefont {Molina-Terriza}},\ }\href@noop {} {\bibfield  {journal}
  {\bibinfo  {journal} {Phys. Rev. Res.}\ }\textbf {\bibinfo {volume} {5}},\
  \bibinfo {pages} {023116} (\bibinfo {year} {2023})}\BibitemShut {NoStop}%
\bibitem [{Note1()}]{Note1}%
  \BibitemOpen
  \bibinfo {note} {In this context, we use the term \protect \emph {photon wave
  function} as a legacy from Bialynicki-Birula's work. We will always deal in
  this work with classical electromagnetic fields, and the photon wave
  functions are electromagnetic wave solutions.}\BibitemShut {Stop}%
\bibitem [{\citenamefont {Bialynicki-Birula}\ and\ \citenamefont
  {Bialynicka-Birula}(2017)}]{IZBirula}%
  \BibitemOpen
  \bibfield  {author} {\bibinfo {author} {\bibfnamefont {I.}~\bibnamefont
  {Bialynicki-Birula}}\ and\ \bibinfo {author} {\bibfnamefont {Z.}~\bibnamefont
  {Bialynicka-Birula}},\ }\href@noop {} {\bibfield  {journal} {\bibinfo
  {journal} {Journal of Optics}\ }\textbf {\bibinfo {volume} {19}},\ \bibinfo
  {pages} {125201} (\bibinfo {year} {2017})}\BibitemShut {NoStop}%
\bibitem [{\citenamefont {Wigner}(1939)}]{Wigner1}%
  \BibitemOpen
  \bibfield  {author} {\bibinfo {author} {\bibfnamefont {E.}~\bibnamefont
  {Wigner}},\ }\href@noop {} {\bibfield  {journal} {\bibinfo  {journal} {Annals
  of Mathematics}\ }\textbf {\bibinfo {volume} {40}},\ \bibinfo {pages} {149}
  (\bibinfo {year} {1939})}\BibitemShut {NoStop}%
\bibitem [{\citenamefont {Bargmann}\ and\ \citenamefont
  {Wigner}(1948)}]{BargmannWigner}%
  \BibitemOpen
  \bibfield  {author} {\bibinfo {author} {\bibfnamefont {V.}~\bibnamefont
  {Bargmann}}\ and\ \bibinfo {author} {\bibfnamefont {E.~P.}\ \bibnamefont
  {Wigner}},\ }\href@noop {} {\bibfield  {journal} {\bibinfo  {journal}
  {Proceedings of the National Academy of Sciences}\ }\textbf {\bibinfo
  {volume} {34}},\ \bibinfo {pages} {211} (\bibinfo {year} {1948})}\BibitemShut
  {NoStop}%
\bibitem [{\citenamefont {Tung}(1985)}]{WuKiTung}%
  \BibitemOpen
  \bibfield  {author} {\bibinfo {author} {\bibfnamefont {W.-K.}\ \bibnamefont
  {Tung}},\ }\href@noop {} {\emph {\bibinfo {title} {Group Theory in
  Physics}}}\ (\bibinfo  {publisher} {World Scientific},\ \bibinfo {year}
  {1985})\BibitemShut {NoStop}%
\bibitem [{\citenamefont {Bargmann}(1947)}]{Bargmann}%
  \BibitemOpen
  \bibfield  {author} {\bibinfo {author} {\bibfnamefont {V.}~\bibnamefont
  {Bargmann}},\ }\href@noop {} {\bibfield  {journal} {\bibinfo  {journal}
  {Annals of Mathematics}\ }\textbf {\bibinfo {volume} {48}},\ \bibinfo {pages}
  {568} (\bibinfo {year} {1947})}\BibitemShut {NoStop}%
\bibitem [{\citenamefont {Lomont}\ and\ \citenamefont
  {Moses}(1962)}]{ZeroMassRep}%
  \BibitemOpen
  \bibfield  {author} {\bibinfo {author} {\bibfnamefont {J.~S.}\ \bibnamefont
  {Lomont}}\ and\ \bibinfo {author} {\bibfnamefont {H.~E.}\ \bibnamefont
  {Moses}},\ }\href@noop {} {\bibfield  {journal} {\bibinfo  {journal} {Journal
  of Mathematical Physics}\ }\textbf {\bibinfo {volume} {3}},\ \bibinfo {pages}
  {405} (\bibinfo {year} {1962})}\BibitemShut {NoStop}%
\bibitem [{\citenamefont {Bacry}(1976)}]{HBacry}%
  \BibitemOpen
  \bibfield  {author} {\bibinfo {author} {\bibfnamefont {H.}~\bibnamefont
  {Bacry}},\ }\href@noop {} {\bibfield  {journal} {\bibinfo  {journal} {Il
  Nuovo Cimento}\ }\textbf {\bibinfo {volume} {32 A}},\ \bibinfo {pages} {448}
  (\bibinfo {year} {1976})}\BibitemShut {NoStop}%
\bibitem [{\citenamefont {Gersten}(1999)}]{Gersten}%
  \BibitemOpen
  \bibfield  {author} {\bibinfo {author} {\bibfnamefont {A.}~\bibnamefont
  {Gersten}},\ }\href@noop {} {\bibfield  {journal} {\bibinfo  {journal}
  {Foundations of Physics Letters}\ }\textbf {\bibinfo {volume} {12}},\
  \bibinfo {pages} {291} (\bibinfo {year} {1999})}\BibitemShut {NoStop}%
\bibitem [{\citenamefont {Dirac}(1936)}]{Dirac}%
  \BibitemOpen
  \bibfield  {author} {\bibinfo {author} {\bibfnamefont {P.~A.~M.}\
  \bibnamefont {Dirac}},\ }\href@noop {} {\bibfield  {journal} {\bibinfo
  {journal} {Proceedings of the Royal Society of London. Series A -
  Mathematical and Physical Sciences}\ }\textbf {\bibinfo {volume} {155}},\
  \bibinfo {pages} {447} (\bibinfo {year} {1936})}\BibitemShut {NoStop}%
\bibitem [{\citenamefont {Landau}\ and\ \citenamefont
  {Lifshitz}(1984)}]{Landau}%
  \BibitemOpen
  \bibfield  {author} {\bibinfo {author} {\bibfnamefont {L.~D.}\ \bibnamefont
  {Landau}}\ and\ \bibinfo {author} {\bibfnamefont {E.~M.}\ \bibnamefont
  {Lifshitz}},\ }\href@noop {} {\emph {\bibinfo {title} {Electrodynamics of
  continuous media}}}\ (\bibinfo  {publisher} {Pergamon Press},\ \bibinfo
  {year} {1984})\BibitemShut {NoStop}%
\bibitem [{\citenamefont {Kong}(1972)}]{BiTheorem}%
  \BibitemOpen
  \bibfield  {author} {\bibinfo {author} {\bibfnamefont {J.~A.}\ \bibnamefont
  {Kong}},\ }\href@noop {} {\bibfield  {journal} {\bibinfo  {journal}
  {Proceedings of the IEEE}\ }\textbf {\bibinfo {volume} {60}},\ \bibinfo
  {pages} {1036} (\bibinfo {year} {1972})}\BibitemShut {NoStop}%
\bibitem [{\citenamefont {Winternitz}(1977)}]{SymmetryBreak1}%
  \BibitemOpen
  \bibfield  {author} {\bibinfo {author} {\bibfnamefont {P.}~\bibnamefont
  {Winternitz}},\ }in\ \href@noop {} {\emph {\bibinfo {booktitle} {Group
  Theoretical Methods in Physics}}},\ \bibinfo {editor} {edited by\ \bibinfo
  {editor} {\bibfnamefont {R.~T.}\ \bibnamefont {Sharp}}\ and\ \bibinfo
  {editor} {\bibfnamefont {B.}~\bibnamefont {Kolman}}}\ (\bibinfo  {publisher}
  {Academic Press},\ \bibinfo {year} {1977})\ pp.\ \bibinfo {pages}
  {549--572}\BibitemShut {NoStop}%
\bibitem [{\citenamefont {Patera}\ \emph
  {et~al.}(1975{\natexlab{a}})\citenamefont {Patera}, \citenamefont
  {Winternitz},\ and\ \citenamefont {Zassenhaus}}]{SymmetryBreak2}%
  \BibitemOpen
  \bibfield  {author} {\bibinfo {author} {\bibfnamefont {J.}~\bibnamefont
  {Patera}}, \bibinfo {author} {\bibfnamefont {P.}~\bibnamefont {Winternitz}},
  \ and\ \bibinfo {author} {\bibfnamefont {H.}~\bibnamefont {Zassenhaus}},\
  }\href@noop {} {\bibfield  {journal} {\bibinfo  {journal} {Journal of
  Mathematical Physics}\ }\textbf {\bibinfo {volume} {16}},\ \bibinfo {pages}
  {1597} (\bibinfo {year} {1975}{\natexlab{a}})}\BibitemShut {NoStop}%
\bibitem [{\citenamefont {Patera}\ \emph
  {et~al.}(1975{\natexlab{b}})\citenamefont {Patera}, \citenamefont
  {Winternitz},\ and\ \citenamefont {Zassenhaus}}]{SBGeneral1}%
  \BibitemOpen
  \bibfield  {author} {\bibinfo {author} {\bibfnamefont {J.}~\bibnamefont
  {Patera}}, \bibinfo {author} {\bibfnamefont {P.}~\bibnamefont {Winternitz}},
  \ and\ \bibinfo {author} {\bibfnamefont {H.}~\bibnamefont {Zassenhaus}},\
  }\href@noop {} {\bibfield  {journal} {\bibinfo  {journal} {Journal of
  Mathematical Physics}\ }\textbf {\bibinfo {volume} {16}},\ \bibinfo {pages}
  {1615} (\bibinfo {year} {1975}{\natexlab{b}})}\BibitemShut {NoStop}%
\bibitem [{\citenamefont {Patera}\ \emph {et~al.}(1977)\citenamefont {Patera},
  \citenamefont {Sharp}, \citenamefont {Winternitz},\ and\ \citenamefont
  {Zassenhaus}}]{SBGeneral2}%
  \BibitemOpen
  \bibfield  {author} {\bibinfo {author} {\bibfnamefont {J.}~\bibnamefont
  {Patera}}, \bibinfo {author} {\bibfnamefont {R.~T.}\ \bibnamefont {Sharp}},
  \bibinfo {author} {\bibfnamefont {P.}~\bibnamefont {Winternitz}}, \ and\
  \bibinfo {author} {\bibfnamefont {H.}~\bibnamefont {Zassenhaus}},\
  }\href@noop {} {\bibfield  {journal} {\bibinfo  {journal} {Journal of
  Mathematical Physics}\ }\textbf {\bibinfo {volume} {18}},\ \bibinfo {pages}
  {2259} (\bibinfo {year} {1977})}\BibitemShut {NoStop}%
\bibitem [{\citenamefont {Gagnon}\ and\ \citenamefont
  {Winternitz}(1988)}]{SBGeneral3}%
  \BibitemOpen
  \bibfield  {author} {\bibinfo {author} {\bibfnamefont {L.}~\bibnamefont
  {Gagnon}}\ and\ \bibinfo {author} {\bibfnamefont {P.}~\bibnamefont
  {Winternitz}},\ }\href@noop {} {\bibfield  {journal} {\bibinfo  {journal}
  {Journal of Physics A: Mathematical and General}\ }\textbf {\bibinfo {volume}
  {21}},\ \bibinfo {pages} {1493} (\bibinfo {year} {1988})}\BibitemShut
  {NoStop}%
\bibitem [{\citenamefont {Beckers}\ \emph {et~al.}(1977)\citenamefont
  {Beckers}, \citenamefont {Patera}, \citenamefont {Perroud},\ and\
  \citenamefont {Winternitz}}]{SymmetryBreak3}%
  \BibitemOpen
  \bibfield  {author} {\bibinfo {author} {\bibfnamefont {J.}~\bibnamefont
  {Beckers}}, \bibinfo {author} {\bibfnamefont {J.}~\bibnamefont {Patera}},
  \bibinfo {author} {\bibfnamefont {M.}~\bibnamefont {Perroud}}, \ and\
  \bibinfo {author} {\bibfnamefont {P.}~\bibnamefont {Winternitz}},\
  }\href@noop {} {\bibfield  {journal} {\bibinfo  {journal} {Journal of
  Mathematical Physics}\ }\textbf {\bibinfo {volume} {18}},\ \bibinfo {pages}
  {72} (\bibinfo {year} {1977})}\BibitemShut {NoStop}%
\bibitem [{\citenamefont {Patera}\ \emph
  {et~al.}(1976{\natexlab{a}})\citenamefont {Patera}, \citenamefont {Sharp},
  \citenamefont {Winternitz},\ and\ \citenamefont
  {Zassenhaus}}]{SymmetryBreak4}%
  \BibitemOpen
  \bibfield  {author} {\bibinfo {author} {\bibfnamefont {J.}~\bibnamefont
  {Patera}}, \bibinfo {author} {\bibfnamefont {R.~T.}\ \bibnamefont {Sharp}},
  \bibinfo {author} {\bibfnamefont {P.}~\bibnamefont {Winternitz}}, \ and\
  \bibinfo {author} {\bibfnamefont {H.}~\bibnamefont {Zassenhaus}},\
  }\href@noop {} {\bibfield  {journal} {\bibinfo  {journal} {Journal of
  Mathematical Physics}\ }\textbf {\bibinfo {volume} {17}},\ \bibinfo {pages}
  {977} (\bibinfo {year} {1976}{\natexlab{a}})}\BibitemShut {NoStop}%
\bibitem [{\citenamefont {Bliokh}\ and\ \citenamefont
  {Nori}(2015)}]{BliokhOperators}%
  \BibitemOpen
  \bibfield  {author} {\bibinfo {author} {\bibfnamefont {K.~Y.}\ \bibnamefont
  {Bliokh}}\ and\ \bibinfo {author} {\bibfnamefont {F.}~\bibnamefont {Nori}},\
  }\href@noop {} {\bibfield  {journal} {\bibinfo  {journal} {Physics Reports}\
  }\textbf {\bibinfo {volume} {592}},\ \bibinfo {pages} {1} (\bibinfo {year}
  {2015})},\ \bibinfo {note} {transverse and longitudinal angular momenta of
  light}\BibitemShut {NoStop}%
\bibitem [{\citenamefont {Fernandez-Corbaton}\ \emph
  {et~al.}(2012)\citenamefont {Fernandez-Corbaton}, \citenamefont
  {Zambrana-Puyalto},\ and\ \citenamefont
  {Molina-Terriza}}]{PhysRevA.86.042103}%
  \BibitemOpen
  \bibfield  {author} {\bibinfo {author} {\bibfnamefont {I.}~\bibnamefont
  {Fernandez-Corbaton}}, \bibinfo {author} {\bibfnamefont {X.}~\bibnamefont
  {Zambrana-Puyalto}}, \ and\ \bibinfo {author} {\bibfnamefont
  {G.}~\bibnamefont {Molina-Terriza}},\ }\href {\doibase
  10.1103/PhysRevA.86.042103} {\bibfield  {journal} {\bibinfo  {journal}
  {Physical Review A}\ }\textbf {\bibinfo {volume} {86}},\ \bibinfo {pages}
  {042103} (\bibinfo {year} {2012})}\BibitemShut {NoStop}%
\bibitem [{\citenamefont {Sarkar}\ and\ \citenamefont
  {Halas}(1997)}]{HalasHilbert}%
  \BibitemOpen
  \bibfield  {author} {\bibinfo {author} {\bibfnamefont {D.}~\bibnamefont
  {Sarkar}}\ and\ \bibinfo {author} {\bibfnamefont {N.~J.}\ \bibnamefont
  {Halas}},\ }\href@noop {} {\bibfield  {journal} {\bibinfo  {journal} {Phys.
  Rev. E}\ }\textbf {\bibinfo {volume} {56}},\ \bibinfo {pages} {1102}
  (\bibinfo {year} {1997})}\BibitemShut {NoStop}%
\bibitem [{\citenamefont {Nienhuis}(2016)}]{PhysRevA_Hilbert}%
  \BibitemOpen
  \bibfield  {author} {\bibinfo {author} {\bibfnamefont {G.}~\bibnamefont
  {Nienhuis}},\ }\href {\doibase 10.1103/PhysRevA.93.023840} {\bibfield
  {journal} {\bibinfo  {journal} {Phys. Rev. A}\ }\textbf {\bibinfo {volume}
  {93}},\ \bibinfo {pages} {023840} (\bibinfo {year} {2016})}\BibitemShut
  {NoStop}%
\bibitem [{\citenamefont {Mackey}(1978)}]{Mackey}%
  \BibitemOpen
  \bibfield  {author} {\bibinfo {author} {\bibfnamefont {G.~W.}\ \bibnamefont
  {Mackey}},\ }\href@noop {} {\emph {\bibinfo {title} {Unitary Group
  Representations in Physics, Probability and Number Theory}}}\ (\bibinfo
  {publisher} {The Benjamin/Cummings Publishing Company, Inc.},\ \bibinfo
  {year} {1978})\BibitemShut {NoStop}%
\bibitem [{\citenamefont {Rossmann}(2002)}]{Rossmann}%
  \BibitemOpen
  \bibfield  {author} {\bibinfo {author} {\bibfnamefont {W.}~\bibnamefont
  {Rossmann}},\ }\href@noop {} {\emph {\bibinfo {title} {Lie groups: An
  introduction through linear groups}}}\ (\bibinfo  {publisher} {Oxford
  University Press},\ \bibinfo {year} {2002})\BibitemShut {NoStop}%
\bibitem [{\citenamefont {Vavilin}\ and\ \citenamefont
  {Fernandez-Corbaton}(2023)}]{PolychromaticTmatrix}%
  \BibitemOpen
  \bibfield  {author} {\bibinfo {author} {\bibfnamefont {M.}~\bibnamefont
  {Vavilin}}\ and\ \bibinfo {author} {\bibfnamefont {I.}~\bibnamefont
  {Fernandez-Corbaton}},\ }\href@noop {} {\enquote {\bibinfo {title} {The
  polychromatic t-matrix},}\ } (\bibinfo {year} {2023}),\ \Eprint
  {http://arxiv.org/abs/2306.07776} {arXiv:2306.07776} \BibitemShut {NoStop}%
\bibitem [{\citenamefont {Kalnins}\ \emph {et~al.}(1975)\citenamefont
  {Kalnins}, \citenamefont {Patera}, \citenamefont {Sharp},\ and\ \citenamefont
  {Winternitz}}]{SymmetryBreak5}%
  \BibitemOpen
  \bibfield  {author} {\bibinfo {author} {\bibfnamefont {E.~G.}\ \bibnamefont
  {Kalnins}}, \bibinfo {author} {\bibfnamefont {J.}~\bibnamefont {Patera}},
  \bibinfo {author} {\bibfnamefont {R.~T.}\ \bibnamefont {Sharp}}, \ and\
  \bibinfo {author} {\bibfnamefont {P.}~\bibnamefont {Winternitz}},\ }in\
  \href@noop {} {\emph {\bibinfo {booktitle} {Group Theory and its
  Applications}}},\ \bibinfo {editor} {edited by\ \bibinfo {editor}
  {\bibfnamefont {E.~M.}\ \bibnamefont {Loebl}}}\ (\bibinfo  {publisher}
  {Academic Press},\ \bibinfo {year} {1975})\ pp.\ \bibinfo {pages}
  {369--464}\BibitemShut {NoStop}%
\bibitem [{\citenamefont {Bialynicki-Birula}(1996)}]{Birula1}%
  \BibitemOpen
  \bibfield  {author} {\bibinfo {author} {\bibfnamefont {I.}~\bibnamefont
  {Bialynicki-Birula}},\ }in\ \href@noop {} {\emph {\bibinfo {booktitle}
  {Progress in Optics}}},\ Vol.~\bibinfo {volume} {36},\ \bibinfo {editor}
  {edited by\ \bibinfo {editor} {\bibfnamefont {E.}~\bibnamefont {Wolf}}}\
  (\bibinfo  {publisher} {Elsevier},\ \bibinfo {year} {1996})\ Chap.~\bibinfo
  {chapter} {5}, pp.\ \bibinfo {pages} {245--294}\BibitemShut {NoStop}%
\bibitem [{\citenamefont {Patera}\ \emph
  {et~al.}(1976{\natexlab{b}})\citenamefont {Patera}, \citenamefont {Sharp},
  \citenamefont {Winternitz},\ and\ \citenamefont
  {Zassenhaus}}]{SubgroupsPoincare}%
  \BibitemOpen
  \bibfield  {author} {\bibinfo {author} {\bibfnamefont {J.}~\bibnamefont
  {Patera}}, \bibinfo {author} {\bibfnamefont {R.~T.}\ \bibnamefont {Sharp}},
  \bibinfo {author} {\bibfnamefont {P.}~\bibnamefont {Winternitz}}, \ and\
  \bibinfo {author} {\bibfnamefont {H.}~\bibnamefont {Zassenhaus}},\
  }\href@noop {} {\bibfield  {journal} {\bibinfo  {journal} {Journal of
  Mathematical Physics}\ }\textbf {\bibinfo {volume} {17}},\ \bibinfo {pages}
  {977} (\bibinfo {year} {1976}{\natexlab{b}})}\BibitemShut {NoStop}%
\bibitem [{\citenamefont {Taylor}(1972)}]{ScatteringTaylor}%
  \BibitemOpen
  \bibfield  {author} {\bibinfo {author} {\bibfnamefont {J.~R.}\ \bibnamefont
  {Taylor}},\ }\href@noop {} {\emph {\bibinfo {title} {Scattering theory: the
  quantum theory of nonrelativistic collisions}}}\ (\bibinfo  {publisher} {John
  Wiley and Sons, Inc.},\ \bibinfo {year} {1972})\BibitemShut {NoStop}%
\bibitem [{\citenamefont {Jackson}(1999)}]{Jackson}%
  \BibitemOpen
  \bibfield  {author} {\bibinfo {author} {\bibfnamefont {J.~D.}\ \bibnamefont
  {Jackson}},\ }\href@noop {} {\emph {\bibinfo {title} {Classical
  Electrodynamics}}}\ (\bibinfo  {publisher} {John Wiley and Sons, Inc.},\
  \bibinfo {year} {1999})\BibitemShut {NoStop}%
\bibitem [{\citenamefont {Fern\'andez-Corbaton}(2014)}]{IvanThesis}%
  \BibitemOpen
  \bibfield  {author} {\bibinfo {author} {\bibfnamefont {I.}~\bibnamefont
  {Fern\'andez-Corbaton}},\ }\href@noop {} {\enquote {\bibinfo {title}
  {Helicity and duality symmetry in light matter interactions: theory and
  applications},}\ } (\bibinfo {year} {2014})\BibitemShut {NoStop}%
\end{thebibliography}%

\end{document}